\documentclass[%
aip,
amsmath,amssymb,
reprint,%
]{revtex4-1}
\usepackage{appendix}
\usepackage{booktabs}    
\usepackage{multirow}    
\usepackage{makecell}    
\usepackage{amsmath}     
\usepackage{braket}      
\usepackage{booktabs,multirow,makecell,amsmath,braket}
\usepackage{xcolor}
\usepackage{enumitem}  
\usepackage[colorlinks,linkcolor=blue,citecolor=blue,urlcolor=blue]{hyperref}
\usepackage{subcaption} 
\usepackage{graphicx}
\usepackage{dcolumn}
\usepackage{bm}
\usepackage[justification=justified]{caption}

\makeatletter
\def\@email#1#2{%
 \endgroup
 \patchcmd{\titleblock@produce}
  {\frontmatter@RRAPformat}
  {\frontmatter@RRAPformat{\produce@RRAP{*#1\href{mailto:#2}{#2}}}\frontmatter@RRAPformat}
  {}{}
}%
\makeatother
\begin{document}

\preprint{AIP/123-QED}

\title{ High-capacity dual degrees of freedom quantum secret sharing protocol beyond the linear rate-distance bound }
\author{Meng-Dong Zhu}
 \affiliation{College of Electronic and Optical Engineering and College of Flexible Electronics (Future Technology), Nanjing University of Posts and Telecommunications, Nanjing, Jiangsu 210023, China}
 \author{Cheng Zhang}
  \affiliation{College of Electronic and Optical Engineering and College of Flexible Electronics (Future Technology), Nanjing University of Posts and Telecommunications, Nanjing, Jiangsu 210023, China}
 \author{Shi-Pu Gu}
 \affiliation{College of Electronic and Optical Engineering and College of Flexible Electronics (Future Technology), Nanjing University of Posts and Telecommunications, Nanjing, Jiangsu 210023, China}
 \author{Xing-Fu Wang}
 \affiliation{College of Science, Nanjing University of Posts and Telecommunications, Nanjing, Jiangsu 210023, China}
  \author{Ming-Ming Du}
 \affiliation{College of Electronic and Optical Engineering and College of Flexible Electronics (Future Technology), Nanjing University of Posts and Telecommunications, Nanjing, Jiangsu 210023, China}
  \author{Wei Zhong}
 \affiliation{Institute of Quantum Information and Technology, Nanjing University of Posts and Telecommunications, Nanjing, Jiangsu 210003, China}
 \author{Lan Zhou}
\altaffiliation{Authors to whom correspondence should be addressed: zhoul@njupt.edu.cn }
\affiliation{School of Physics, Hangzhou Normal University, Hangzhou, Zhejiang 311121, China}
\author{Yu-Bo Sheng}
\altaffiliation{Authors to whom correspondence should be addressed: shengyb@njupt.edu.cn }
\affiliation{College of Electronic and Optical Engineering and College of Flexible Electronics (Future Technology), Nanjing University of Posts and Telecommunications, Nanjing, Jiangsu 210023, China}

\date{\today}

\begin{abstract}
Quantum secret sharing (QSS) is the multipartite cryptographic primitive. Most of existing QSS  protocols are limited by the linear rate-distance bound, and cannot realize the long-distance and high-capacity multipartite key distribution. This paper proposes a polarization (Pol) and phase (Ph) dual degrees of freedom (dual-DOF) QSS protocol based on the weak coherent pulse (WCP) sources. Our protocol combines the single-photon interference, two-photon interference and non-interference principles, and can resist the internal attack from the dishonest player. We develop simulation method to estimate its performance under the beam splitting attack. The simulation results show that our protocol can surpass the linear bound. Comparing with the differential-phase-shift twin-field QSS and WCP-Ph-QSS protocols, our protocol has stronger resistance against the beam splitting attack, and thus has longer maximal communication distance and higher key rate.  By using the WCPs with high average photon number ($\mu=1.5$), our protocol achieves a key rate about 5.4 times of that in WCP-Ph-QSS protocol. Its maximal communication distance (441.7 km) is about 7.9\% longer than that of the WCP-Ph-QSS. Our protocol is highly feasible with current experimental technology and offers a promising approach for long-distance and high-capacity quantum networks.
\end{abstract}

\maketitle
\section{Introduction}
Quantum communication guarantees the security of information transmission based on the fundamental principles of quantum mechanics. Quantum communication encompasses typical branches including quantum key distribution (QKD) \cite{qkd1,qkd2,qkd3,qkd5,qkd6,qkd7,qkd8}, quantum secret sharing (QSS) \cite{qss_first,qss1,qss2,qss3} and quantum secure direct communication \cite{qsdc1,qsdc2,qsdc3,qsdc4}. QKD enables two remote users to generate secret keys. QSS is a critical multipartite cryptographic primitive, which enables multiple players to decode the keys from a dealer by cooperation. QSS has been regarded as a critical part of the future quantum network.
The first QSS protocol based on the Greenberger-Horne-Zeilinger (GHZ) states was proposed in 1999 \cite{qss_first}. Since then, QSS has been widely researched in theory \cite{qss5,qss6,qss6n,qss7,qss9,qss10,qss11,qss12,diqss,diqss1,diqss2,mdiqss2,qss_single_qubit1n,qss_single_qubit2,qss_single_qubit3}. On one hand, some QSS protocols replace the GHZ state with the Bell states \cite{qss5} and single-qubits \cite{qss_single_qubit1n,qss_single_qubit2} to reduce its experimental difficulty. On the other hand, the high-security QSS protocols such as the device-independent QSS \cite{diqss,diqss1,diqss2} and measurement-device-independent (MDI) QSS protocols \cite{qss6n,mdiqss2} have been proposed. QSS also has achieved great experimental progresses \cite{qss_experiment1,QSSe3,QSSe4,qss_experiment4,qss_experiment5,qss_experiment2,QSSe8}.

Similar as QKD, the key rate of QSS decreases exponentially with the transmission distance due to the environmental noise. The key rate is limited by the
linear rate-distance bound \cite{linear_rate1,linear_rate2n,linear_rate3,linear_rate4,linear_rate3n}.
Since 2018, the twin-field (TF) QKD protocols based on the single-photon interference have been put forward \cite{TF-QKD1,TF-QKD2,TF-QKD3,TF-QKD4}, which builds a promising key rate-distance relationship to surpass the linear rate-distance bound.
Later, the single-photon interference principle was also adopted in QSS \cite{WCP_Ph_QSS,DPS_TF_QSS,RR_QSS}. In 2021, the differential-phase-shift (DPS) TF QSS based on the weak coherent pulse (WCP) sources  was proposed \cite{DPS_TF_QSS}.
In 2023, Shen \emph{et al.} proposed the QSS protocol based on the phase-encoded WCPs, say, the WCP-Ph-QSS protocol \cite{WCP_Ph_QSS}.
    A feasible approach to increase QSS's key rate is the utilization of high-dimensional quantum states \cite{high-dimension,high-dimension1}. However, there have been lack of the research on the high-dimensional long-distance QSS protocols. In 2023, the interfering-or-not-interfering (INI) QKD protocol first adopted the polarization-phase hyperencoding technology to achieve the higher key rate \cite{INI_qkd}.

 Inspired by the WCP-Ph-QSS \cite{WCP_Ph_QSS} and INI-QKD \cite{INI_qkd} protocols, in the paper, we propose the first high-capacity dual-DOF QSS protocol based on the WCP sources. Two players construct the polarization-phase dual degrees of freedom (dual-DOF) bases to encode the WCPs. They send the encoded WCPs to the dealer for the Bell state measurement (BSM). Our protocol combines the single-photon interference, two-photon interference and non-interference principles. We develop simulation method to estimate its key rate under the beam splitting attack \cite{BSA1,BSA2}. The results show that our protocol achieves the key rate beyond the linear rate-distance bound and has stronger resistance against the beam splitting attack than the WCP-Ph-QSS and DPS-TF-QSS protocols.
   With the average photon number $\mu=1.5$, our protocol has a key rate of $1.99\times10^{-6}$ bit/pulse with $L=400$ km, about 5.4 times of that in WCP-Ph-QSS protocol. Its maximal communication distance (441.7 km) is 7.9\% longer than that of the WCP-Ph-QSS protocol.
Our QSS protocol offers a promising approach for realizing the high-capacity and long-distance quantum networks in the future.

\section{\label{sec.2}The dual-DOF QSS protocol under the beam splitting attack}
\begin{figure}[htbp]
\centering
\includegraphics[width=8.5cm]{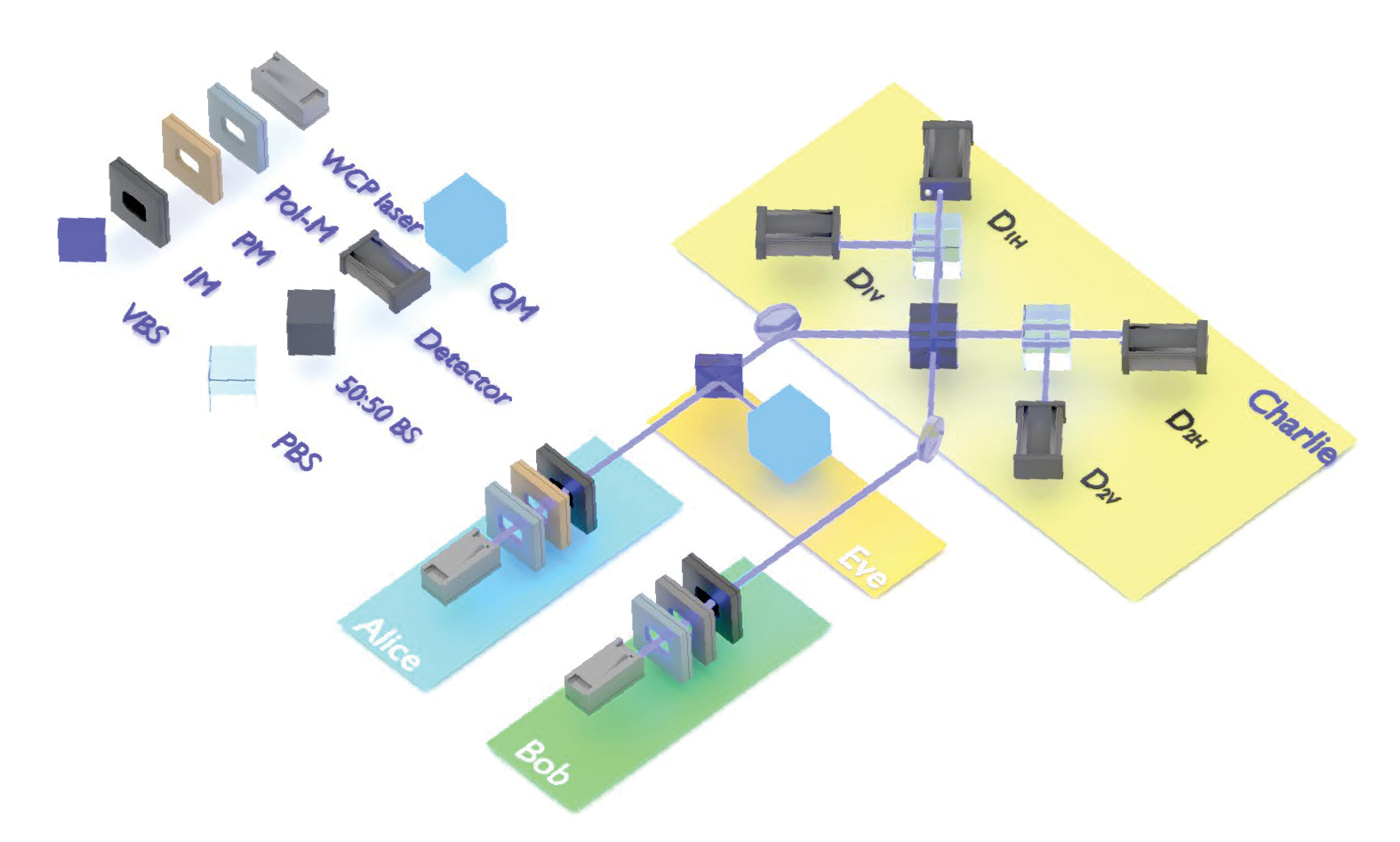}
\caption{Schematic diagram of the dual-DOF QSS protocol under the beam splitting attack. The WCPs are generated from the laser sources. The polarization modulator (Pol-M) and phase modulator (PM) are used to realize the polarization and phase encoding, respectively, and the intensity modulator (IM) realizes the intensity modulation. PBS, BS and VBS represent polarization beam splitter, 50:50 beam splitter and variable beam splitter. QM represents quantum memory. $D_{1H}$, $D_{1V}$, $D_{2H}$, $D_{2V}$ are practical photon detectors.}
\label{FIG.1}
\end{figure}

 In our protocol, Charlie is the dealer while Alice and Bob are two players. Its basic diagram under the beam splitting attack is shown in Fig. \ref{FIG.1}. The protocol includes six steps as follows.

Step 1. Alice and Bob generate the WCPs with the average photon number of $\mu$ from the WCP sources. They each randomly encode the pulse in rectilinear ($Z_{dual}$) or diagonal ($X_{dual}$) dual-DOF basis with the form of
\begin{eqnarray}
Z_{dualA(B)}&=\{|\sqrt\mu\rangle_{{A(B)}^H},|\sqrt\mu \rangle_{{A(B)}^V}, \nonumber\\
&|-\sqrt\mu\rangle_{{A(B)}^H},|-\sqrt\mu \rangle_{{A(B)}^V}\},\nonumber\\
X_{dualA(B)}&=\{\left.\left|\sqrt\mu \right.\right\rangle_{{A(B)}^+},\left.\left|\sqrt\mu \right.\right\rangle_{{A(B)}^-},\nonumber\\
&\left.\left|-\sqrt\mu \right.\right\rangle_{{A(B)}^+},\left.\left|-\sqrt\mu \right.\right\rangle_{{A(B)}^-}\},
\end{eqnarray}
 where $|H\rangle$ ($|V\rangle$) represents the horizontal (vertical) polarization, $|+\rangle=\frac{1}{\sqrt{2}}(|H\rangle+|V\rangle)$ and $|-\rangle=\frac{1}{\sqrt{2}}(|H\rangle-|V\rangle)$, and A (B) represents Alice (Bob).

The polarization bit and phase bit are denoted as $\kappa_{A(B)}^{Pol}\in\{0, 1\}$ and $\kappa_{A(B)}^{Ph}\in\{0, 1\}$, respectively. The key bits $\kappa^{Pol}\kappa^{Ph}=00$ are encoded as $\left.\left|\sqrt\mu \right.\right\rangle_{{A\left(B\right)}^+}$ or $\left.\left|\sqrt\mu\right.\right\rangle_{{A\left(B\right)}^H}$, 01 are encoded as $\left.\left|-\sqrt\mu \right.\right\rangle_{{A\left(B\right)}^+}$ or $\left.\left|-\sqrt\mu\right.\right\rangle_{{A\left(B\right)}^H}$, 10 are encoded as $\left.\left|\sqrt\mu \right.\right\rangle_{{A\left(B\right)}^-}$ or $\left.\left|\sqrt\mu\right.\right\rangle_{{A\left(B\right)}^V}$, and 11 are encoded as $\left.\left|-\sqrt\mu \right.\right\rangle_{{A\left(B\right)}^-}$ or $\left.\left|-\sqrt\mu\right.\right\rangle_{{A\left(B\right)}^V}$.

Step 2. Alice and Bob send the encoded pulses to Charlie located at an intermediate station between them through quantum channels. The external eavesdropper (Eve) performs the beam splitting attack. He uses a VBS on the channel from Alice to Charlie. The reflected pulses are stored in his QM and the transmitted pulses are sent to Charlie through the ideal lossless quantum channel.

Step 3. Charlie performs the BSM on received pulses and records the detector responses. Three effective response events are denoted as Event$_1$, Event$_2$ and Event$_3$, respectively, where
\begin{enumerate}[
    label=(\arabic*),
    leftmargin=1.5em,
    itemindent=*,
    itemsep=0pt,
    parsep=4pt
]
\item Event$_1$: only the photon detector $D_{1H}$ or $D_{2H}$ responds.
\item Event$_2$: the photon detectors $(D_{1H}, D_{1V})$ or $(D_{2H}, D_{2V})$ respond.
\item Event$_3$: the photon detectors $(D_{1H}, D_{2V})$ or $(D_{1V}, D_{2H})$ respond.
\end{enumerate}
  The other detector responses are treated as failures. Charlie only announces the locations corresponding to the single-click events (Y$_1$, equals to Event$_1$) and double-click events (Y$_2$, including Event$_2$ and Event$_3$). Alice and Bob preserve the encoded bits corresponding to Y$_1$ and Y$_2$.

   In Y$_1$, Charlie only extracts $\kappa_{C}^{Ph}$ from the detector responses, which will be 0 (1) when the detector $D_{1H}$ $(D_{2H})$ clicks. In Y$_2$, Charlie can extract $\kappa_{C}^{Ph}\kappa_{C}^{Pol}$ from the detector responses. In detail, $\kappa_{C}^{Ph}\kappa_{C}^{Pol}$ will be 00 when $(D_{1H}, D_{1V})$ respond and 10 when ($D_{2H}$, $D_{2V}$) respond. $\kappa_{C}^{Ph}\kappa_{C}^{Pol}$ will be 01 when ($D_{1H}$, $D_{2V}$) respond and 11 when ($D_{2H}$, $D_{1V}$) respond.

Step 4. The above processes are repeated until Charlie obtains enough key bits.

Step 5. Alice and Bob announce the generation basis $Z_{dual}$ or $X_{dual}$ for each of their preserved bits and discard the bits corresponding to different generation bases. Next, Bob announces all the preserved bits in $Z_{dual}$ basis and $X_{dual}$ basis, and Alice announces all of her preserved bits in $Z_{dual}$ basis and randomly announces a part of preserved bits in $X_{dual}$ basis for security checking. Charlie uses the announced bits and his corresponding BSM results to estimate the quantum bit error rate (QBER). Alice's remained bits and Bob's corresponding bits in $X_{dual}$ basis form their raw key bits. Charlie's corresponding bits form his raw key bits. We provide the relationship among the three users' bits in Y$_1$ and Y$_2$  in Tab. \ref{tab:1}. The corresponding thresholds of the QBERs in $Z_{dual}$ ($X_{dual}$) in Y$_1$ event and Y$_2$ event are calculated as 2.39\% and 2.08\%, respectively.
If the practical QBER in Y$_1$ or Y$_2$ is higher than the tolerable threshold, the users have to discard their key bits. On the contrary, if the practical QBERs in both Y$_1$ and Y$_2$  are below the tolerable thresholds, the users ensure the security of key generation process and preserve their raw key bits.
\begin{table}[t]
    \centering
    \vspace{-0.1cm}
    \setlength{\abovecaptionskip}{0.3cm}
    \setlength{\belowcaptionskip}{0.1cm}
    \setlength\tabcolsep{8pt}
    \renewcommand\arraystretch{1.5}
    \caption{The relationship among the three users' bits in Y$_1$ and Y$_2$.}
    \begin{tabular}{ccccc}
        \hline \hline
        Event&Alice&Bob&Charlie\\
       \hline
       Y$_1$ & $\kappa_a^{Ph}$ &$\kappa_b^{Ph}$ &$\kappa_a^{Ph}\oplus\kappa_b^{Ph}$ \\
       Y$_2$&$\kappa_a^{Ph}\kappa_a^{Pol}$ &$\kappa_b^{Ph}\kappa_b^{Pol}$ &$\kappa_a^{Ph}\oplus\kappa_b^{Ph},\kappa_a^{Pol}\oplus\kappa_b^{Pol}$\\
        \hline
        \hline
        \label{tab:1}
    \end{tabular}
\end{table}

Step 6. Three users perform the classical error correction and privacy amplification on their raw key bits to distill the secure key bits.

Step 7. Bob announces his secure key bits and Alice can reconstruct the key bits delivered by Charlie.

\section{\label{sec.3}Security analysis}
As our protocol combines the single-photon interference, two-photon interference and non-interference principles based on the WCP sources,
 its security proof can refer to those of the WCP-Ph-QSS \cite{WCP_Ph_QSS} and the INI-QKD \cite{INI_qkd}. Then, we analyze its security against the internal attack from the dishonest Bob and external beam splitting attack.

\textbf{The internal attack.} We assume that Bob is a dishonest user who falsely declares the generated basis or encoded bits. As Alice's encoded bits and Charlie's detector results are private, Bob cannot obtain Alice's or Charlie's raw key bits. However, Bob's attack would make Alice deduce wrong key bits from Charlie.
Fortunately, this internal attack  can be detected by the security checking. In detail, Bob has to announce all the encoded bits in $Z_{dual}$ and $X_{dual}$ bases and he does not know the positions of the security checking bits in $X_{dual}$ basis. In this way, his false announcement inevitably increases the QBER  and thus be detected in the security checking. As a result, our QSS protocol can resist the internal attack from Bob.

\textbf{The external attack.} Eve performs the beam splitting attack as shown in Step 2. As the transmittance of the VBS precisely matches the photon transmission efficiency $\eta_t$ of the actual quantum channel and the eavesdropping does not change the encoded bits of the pulses, this attack will not influence the gains in Charlie's photon detectors or increase the QBER of the security checking.
 Subsequently, after Alice announcing the generation basis for each WCP and the locations of the key encoded WCPs, Eve can extract the corresponding photons to make the unambiguous state discrimination (USD) measurement and finally extract Alice's key bits. As all of Bob's key bits should be announced, Eve does not require to attack Bob's quantum channel. Combined with Bob's corresponding key bits, Eve can finally deduce Charlie's key bit. The beam splitting attack can not be detected by the users, so that it has become the primary attack model in the WCP-based quantum communication \cite{DPS_TF_QSS,INI_qkd,BSA1,BSA2}.

\section{\label{sec.4}Final key generation rate}
 As the beam splitting attack is a general attack for Event$_1$, Event$_2$ and Event$_3$, we assume that Alice's key leakage rate ($I_E$) to Eve in above three events are identical. Based on WCP-Ph-QSS \cite{WCP_Ph_QSS} and INI-QKD \cite{INI_qkd} protocols, we can obtain the key rate of our dual-DOF QSS protocol as
\begin{equation}\label{RR}
R=\sum_{i=1}^{3}{Q_{Event_i}[1-I_E-H\left(E_{Event_i}^{ph}\right)-fH\left(E_{Event_i}^{bit}\right)]}.
\end{equation}
Here, $Q_{Event_i}$ denotes the overall gain in Event$_i$ $(i \in \{1, 2, 3\})$, $f$ denotes the error-correction efficiency. $E_{Event_i}^{ph}$ and $E_{Event_i}^{bit}$ denote the phase-error rate and bit-error rate of Charlie's BSM results in Event$_i$, respectively, and $H(x)=- xlog_2x - (1-x) log_2(1 - x)$ is the binary Shannon entropy function.

 Eve's successful unambiguous identification probability of the intercepted pulse can be upper bounded by
 \begin{eqnarray}
P_{\mathrm{success}}&\leqslant&1-\frac{1}{N-1}\sum_{i\neq j}\sqrt{p_ip_j}|\langle\psi_i\mid\psi_j\rangle|,\nonumber\\
&=&1-\frac{1}{3}[e^{-2(1-\eta_{t})\mu}+2e^{-(1-\eta_{t})\mu}],
\end{eqnarray}
where $|\psi_{i(j)}\rangle$ ($i,j=1,2,3,4$) belongs to the four possible quantum states in  $X_{dual}$ basis.
Eve can obtain Alice's key bit when he can successfully identify the quantum states. In this way, we can obtain $I_E=P_{\mathrm{success}}$.

Then, we estimate $Q_{Event_i}$, $E_{Event_i}^{bit}$ and $E_{Event_i}^{ph}$ ($i=1,2,3$) in practical imperfect experiment. $Q_{Event_i}$ can be easily obtained from the amount of probabilities of all possible detector responses in Event$_i$. It is noted that the imperfect photon detectors can induce dark counts, and may cause  bit-flip and phase-flip errors of the BSM. Without loss of generality, we focus on specific cases where Alice's and Bob's phase-encoded bits are $(0, 0)$, and polarization-encoded bits are (0, 0) and (0, 1) to exemplify the general scenario due to the symmetry \cite{INI_qkd}. The wrong detector responses caused by the bit-flip error and the correct detector responses under these specific two scenarios are shown in Tab. \ref{table2}. It is noted that in Event$_1$, only the phase-encoded bits can be extracted from the correct detector response, while the polarization-encoded bit has no correlation with the detector response.

\begin{table}[htbp]
\caption{The correct detector responses ($"\checkmark"$) and wrong detector responses ($"\times"$) caused by the bit-flip error in $\kappa_{a}^{Ph} \kappa_{b}^{Ph} \kappa_{a}^{Pol} \kappa_{b}^{Pol}=(0000)$ and (0001) scenarios. $"\circ"$ means that the encoded bits has no correlation from the detector response. }
\setlength{\arrayrulewidth}{0.3pt}
\setlength\tabcolsep{1pt}
\renewcommand{\arraystretch}{1.3} 
\begin{tabular}{c|cc|cc|cc}
	\hline
	\hline
	\multirow{2}{*}{\makecell{bit correlation\\$(\kappa_{a}^{Ph} \kappa_{b}^{Ph} \kappa_{a}^{Pol} \kappa_{b}^{Pol})$}}
	& \multicolumn{2}{c|}{Event$_1$} & \multicolumn{2}{c|}{Event$_2$} & \multicolumn{2}{c}{Event$_3$} \\
	\cline{2-7}
	& \scalebox{0.8}{$D_{1H}$} & \scalebox{0.8}{$D_{2H}$} & \scalebox{0.8}{$D_{1H},D_{1V}$} & \scalebox{0.8}{$D_{2H},D_{2V}$} & \scalebox{0.8}{$D_{1H},D_{2V}$} & \scalebox{0.8}{$D_{2H},D_{1V}$} \\
	\hline
	$Pol(0000)$ & \scalebox{1.6}{$\circ$} & \scalebox{1.6}{$\circ$} & $\checkmark$ & $\checkmark$ & $\times$ & $\times$ \\
	$Ph(0000)$ & $\checkmark$ & $\times$ & $\checkmark$ & $\times$ & $\checkmark$ & $\times$ \\
	$Pol(0001)$ & \scalebox{1.6}{$\circ$} & \scalebox{1.6}{$\circ$} & $\times$ & $\times$ & $\checkmark$ & $\checkmark$ \\
	$Ph(0001)$ & $\checkmark$ & $\times$ & $\checkmark$ & $\times$ & $\checkmark$ & $\times$ \\
	\hline
	\hline
\end{tabular}
\label{table2}
\end{table}

\begin{table*}[t]
	\centering
	\caption{The error rate of each detector response in $\kappa_{a}^{Ph} \kappa_{b}^{Ph} \kappa_{a}^{Pol} \kappa_{b}^{Pol}=(0000)$ and (0001) scenarios under the bit-flip and phase-flip errors.}
	\setlength{\tabcolsep}{0.5pt}
	\setlength{\arrayrulewidth}{0.4pt}
	\renewcommand{\arraystretch}{1.5}
	\begin{tabular}{c|cc|cc|cc}
		\hline\hline
		\multicolumn{1}{c|}{\multirow{2}{*}{\makecell{bit correlation\\($\kappa_{a}^{Ph} \kappa_{b}^{Ph} \kappa_{a}^{Pol} \kappa_{b}^{Pol})$}}} & \multicolumn{2}{c|}{Event$_1$} & \multicolumn{2}{c|}{Event$_2$} & \multicolumn{2}{c}{Event$_3$} \\
		\cline{2-7}
		\multicolumn{1}{c|}{} & $D_{1H}$ & $D_{2H}$ & $D_{1H},D_{1V}$ & $D_{2H},D_{2V}$ & $D_{1H},D_{2V}$ & $D_{2H},D_{1V}$ \\
		\hline
		$Pol(0000)$ &\scalebox{1.6}{$\circ$}& \scalebox{1.6}{$\circ$} & \resizebox{0.85\width}{!}{$P_r^{H_1,V_1}(o,e)+P_r^{H_1,V_1}(e,e)$}&  \resizebox{0.85\width}{!}{$P_r^{H_1,V_1}(o,e)+P_r^{H_1,V_1}(e,e)$} & \resizebox{0.85\width}{!}{$P_r(D_{1H}D_{2V}|++)$} & \resizebox{0.85\width}{!}{$P_r(D_{1V}D_{2H}|++)$} \\[0.6ex]
		$Ph(0000)$ & $P_r^{H_1}(e)$ & \resizebox{0.85\width}{!}{$P_r(D_{2H}|++)$} & \resizebox{0.85\width}{!}{$P_r^{H_1,V_1}(o,e)+P_r^{H_1,V_1}(e,o)$}&   \resizebox{0.85\width}{!}{$P_r(D_{2H}D_{2V}|++)$} & \resizebox{0.85\width}{!}{$P_r^{H_1,V_1}(o,e)+P_r^{H_1,V_1}(e,o)$} & \resizebox{0.85\width}{!}{$P_r(D_{1V}D_{2H}|++)$} \\[0.6ex]
		$Pol(0001)$ & \scalebox{1.6}{$\circ$} & \scalebox{1.6}{$\circ$} & \resizebox{0.85\width}{!}{$P_r(D_{1H}D_{1V}|+-)$} & \resizebox{0.85\width}{!}{$P_r(D_{2H}D_{2V}|+-)$} & \resizebox{0.85\width}{!}{$P_r^{H_1,V_2}(o,e)+P_r^{H_1,V_2}(e,e)$} & \resizebox{0.85\width}{!}{$P_r^{H_2,V_1}(o,e)+P_r^{H_2,V_1}(e,e)$} \\[0.6ex]
		$Ph(0001)$ & $P_r^{H_1}(e)$ & \resizebox{0.85\width}{!}{$P_r(D_{2H}|+-)$} & \resizebox{0.85\width}{!}{$P_r^{H_1,V_1}(o,e)+P_r^{H_1,V_1}(e,0)$} & \resizebox{0.85\width}{!}{$P_r(D_{2H}D_{2V}|+-)$} & \resizebox{0.85\width}{!}{$P_r^{H_1,V_2}(o,e)+P_r^{H_1,V_2}(e,o)$} & \resizebox{0.85\width}{!}{$P_r(D_{1V}D_{2H}|+-)$} \\
		\hline\hline
	\end{tabular}
	\label{table3}
\end{table*}

The correct detector response to incident even photon numbers would lead to the phase-flip error in BSM.
Considering both the bit-flip and phase-flip errors, we list the error rate of each detector response under the (0000) and (0001) scenarios in Tab. \ref{table3}.
Here, $P_r$ represents the probability, while $o$ and $e$ mean the odd and even photon number in the WCPs, respectively.
$E_{Event_i}^{bit}$ ($i=1,2,3$) equals to the amount of the error probabilities in Tab. \ref{table3} of the items with $"\times"$ in the $Event_i$ column of Tab. \ref{table2}, while $E_{Event_i}^{ph}$ ($i=1,2,3$) equals to the amount of the error probabilities in Tab. \ref{table3} of the items with $"\checkmark"$ in the $Event_i$ column of Tab. \ref{table2}. By taking the values of $Q_{Event_i}$, $E_{Event_i}^{bit}$, $E_{Event_i}^{ph}$ ($i=1,2,3$) and $I_E$ into Eq. (\ref{RR}), we can obtain the lower bound of $R$.

\section{\label{sec.5}Numerical simulation}
In this section, we suppose that the total communication distance between Alice and Bob is $L$, so that the photon transmission distance from Alice (Bob) to Charlie is $\frac{L}{2}$. The transmittance $\eta_t=\eta_d 10^{-\alpha L/20}$, where $\eta_d$ is detection efficiency of Charlie's detectors.

\begin{figure}[htbp]
\centering
\includegraphics[scale=0.28]{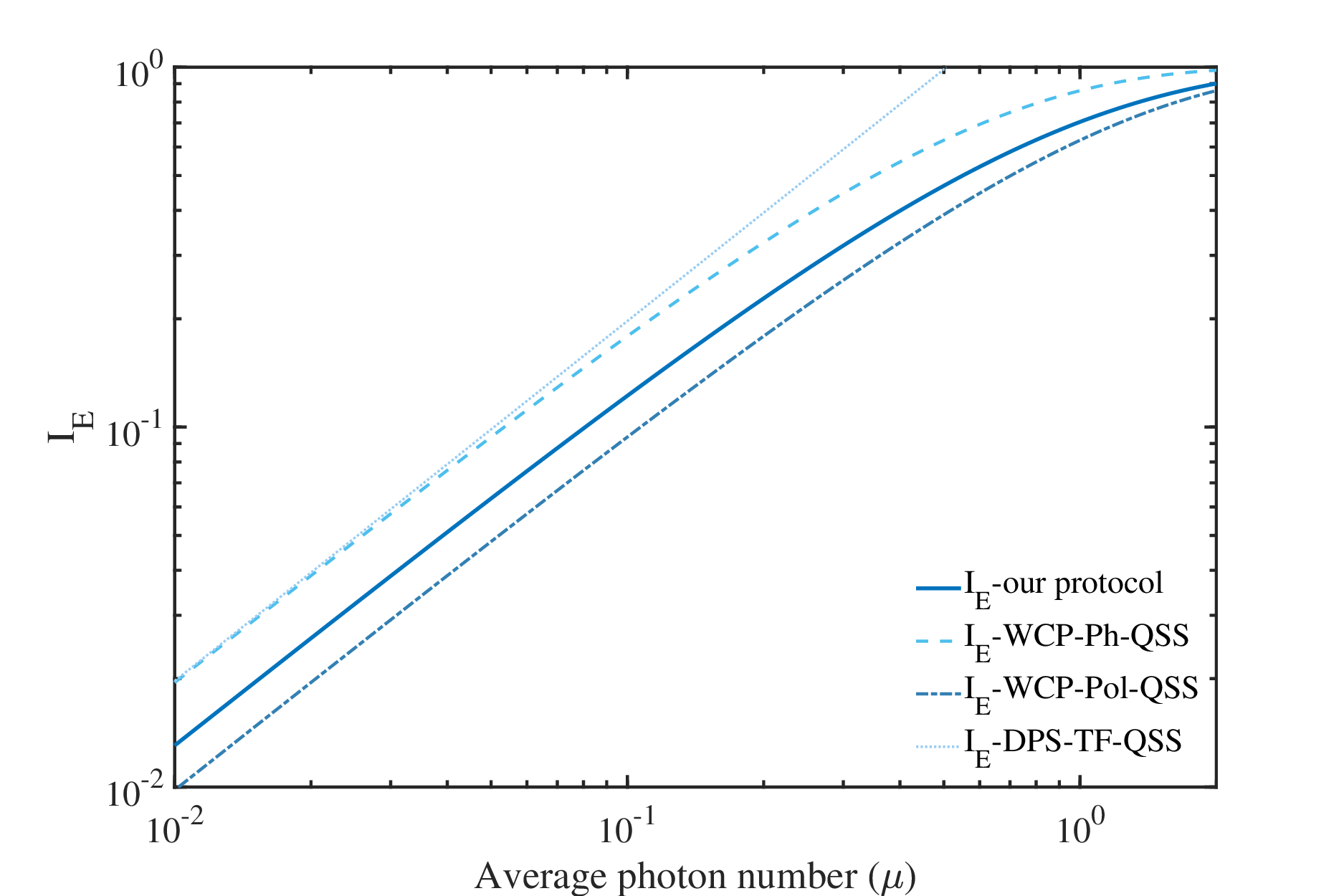}
\caption{$I_E$ of our dual-DOF QSS protocol, WCP-Ph-QSS protocol \cite{WCP_Ph_QSS}, WCP-Pol-QSS protocol and DPS-TF-QSS protocol \cite{DPS_TF_QSS} under the beam splitting attack altered with $\mu$ at $L=100$ km. $I_E$ of the WCP-Pol-QSS protocol equals to that of the MDI-QKD protocol \cite{qkd5}.}
\label{FIG.2.0}
\end{figure}

As shown in Fig. \ref{FIG.2.0},  $I_E$ of the four QSS protocols all increase with the growth of $\mu$. The WCP-Pol-QSS protocol has the lowest $I_E$. The reason is that it only uses the two-photon interference to generate keys, which has the lowest probability for Eve obtaining Charlie's key only from Alice's key bit. All the other three protocols require the single-photon interference in phase DOF, which are relatively fragile to the beam splitting attack.
Our protocol has lower $I_E$ than the WCP-Ph-QSS \cite{WCP_Ph_QSS} and DPS-TF-QSS protocols \cite{DPS_TF_QSS}.
When $\mu=0.4$, the values of $I_E$ in the DPS-TF-QSS protocol, WCP-Ph-QSS protocol and our protocol are about 0.789, 0.545, and 0.399, respectively. The DPS-TF-QSS protocol has the threshold of $\mu=0.5074$, for it only uses the single-photon interference for key generation and high value of $\mu$ will amplify its security vulnerabilities from mult-photon pulses.
Our protocol and the WCP-Ph-QSS also use the multi-photon pulse components, so that it can tolerate higher $\mu$. Moreover, as the WCP-Ph-QSS protocol only has two possible states in each basis \cite{WCP_Ph_QSS} but our protocol has four, our protocol can reduce Eve's probability of correctly guessing the result, thus further reduce $I_E$.

\begin{figure}[htbp]
\centering
\includegraphics[scale=0.55]{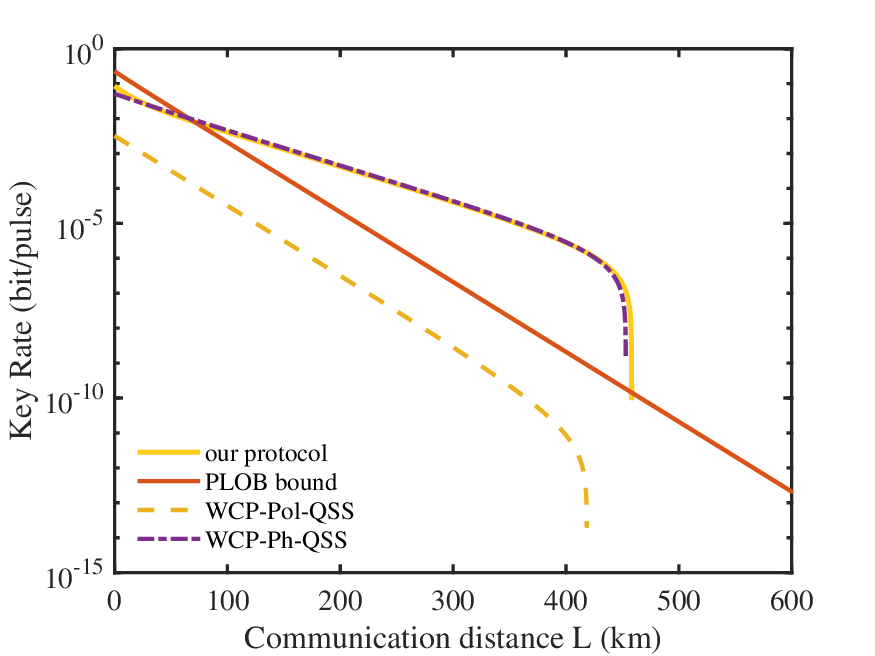}
\caption{The key rates of our dual-DOF QSS, WCP-Ph-QSS \cite{WCP_Ph_QSS} and WCP-Pol-QSS \cite{qkd5} altered with the communication distance ($\mu=0.84$). The red line represents the PLOB bound \cite{linear_rate3}}
\label{FIG.4}
\end{figure}

\begin{figure}[htbp]
\centering
\includegraphics[scale=0.55]{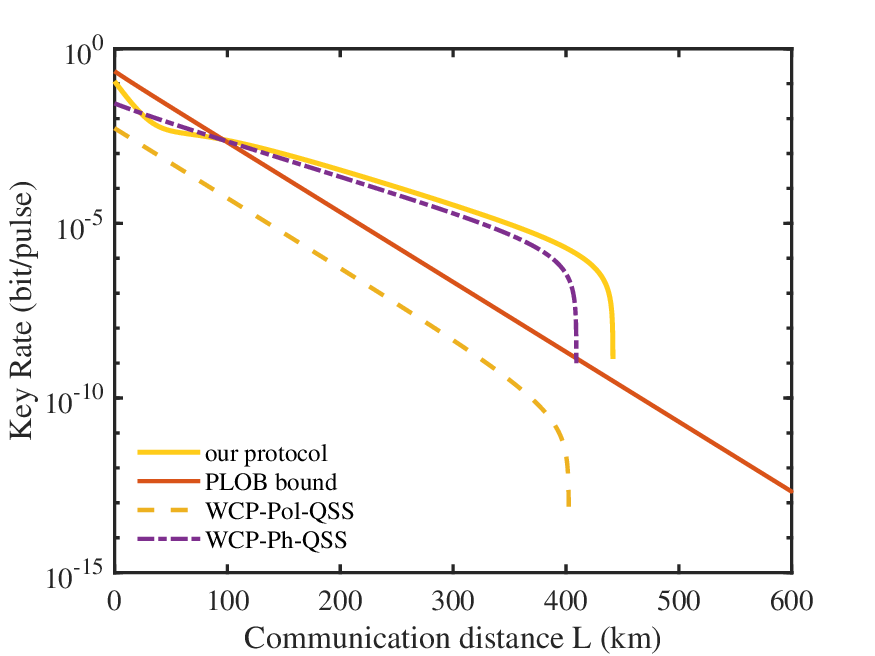}
\caption{The key rates of our dual-DOF QSS, WCP-Ph-QSS \cite{WCP_Ph_QSS} and WCP-Pol-QSS \cite{qkd5} altered with the communication distance ($\mu=1.5$). The red line represents the PLOB bound \cite{linear_rate3}}
\label{FIG.5}
\end{figure}

We optimize the value of $\mu$ by the genetic algorithm to maximize the key generation rate of our protocol.
We fix $L=400$ km and obtain the approximate optimal value of $\mu=0.84$. In Fig. \ref{FIG.4} and Fig. \ref{FIG.5}, we compare the key rates of our dual-DOF QSS with the WCP-Ph-QSS  \cite{WCP_Ph_QSS} and the WCP-Pol-QSS with $\mu=0.84$ and $\mu=1.5$, respectively (DPS-TF-QSS does not have positive key rate in these cases). Here, we set $\eta_d=0.145$ \cite{INI_qkd,WCP_Ph_QSS,DPS_TF_QSS}, the attenuation coefficient of the fiber $\alpha=0.2$ dm/km, the dark count rate $P_d=8 \times 10^{-8}$, and the error correction efficiency $f= 1.15$. The key rate of the WCP-Pol-QSS equals to that of the MDI-QKD \cite{qkd5}. WCP-Pol-QSS protocol cannot break through the linear bound and contributes lower secret key rates as it depends on the two-photon interference.
Both the WCP-Ph-QSS and our QSS protocols achieve key rates beyond the linear rate-distance bound benefit to the single-photon interference.
 The WCP-Ph-QSS protocol displays a slightly higher key rate under $\mu=0.84$.
 When $L=400$ km, the WCP-Ph-QSS protocol and our protocol have the key rate of $2.91\times 10^{-6}$ and $2.83\times 10^{-6}$, respectively. However, due to the lower $I_E$, our protocol has a longer maximal communication distance $L$. The maximal values of $L$ are 453.1 km and  458.3 km for the WCP-Ph-QSS protocol and our protocol, respectively. The advantage in $I_E$ of our protocol becomes more obvious with the growth of $\mu$. When $\mu=1.5$, our protocol has the key rate $1.99\times10^{-6}$ bit/pulse under $L=400$ km, which is about 5.4 times of that in the WCP-Ph-QSS protocol ($3.66\times10^{-7}$ bit/pulse). Its maximal communication distance (441.7 km) is about 7.9\% longer than those of the WCP-Ph-QSS protocol (409.3 km). It indicates that our dual-DOF QSS protocol exhibits stronger advantages than the WCP-Ph-QSS protocol in high $\mu$ scenario.

\section{conclusion}
QSS is a critical multipartite cryptographic primitive, which enables multiple players to decode keys from a dealer by cooperation. Most of existing QSS  protocols are limited by the linear rate-distance bound, and thus cannot realize the long-distance and high-capacity multipartite key distribution. In this work, we propose a dual-DOF QSS protocol based on the WCP sources. Two players randomly use the polarization-phase dual-DOF bases to generate WCPs. They send the encoded WCPs to the dealer for the measurement. Based on the detector responses, the dealer may obtain one bit of key from single-photon interference, or two bits of keys from the two-photon interference or non-interference per measurement round. Our protocol can resist the internal attack. We develop simulation method to estimate its key rates under the beam splitting attack. Our protocol can achieve the key rate beyond the linear rate-distance bound due to the single-photon interference component. We obtain the approximate optimal value of the average number $\mu=0.84$, which makes our protocol achieve the maximal communication distance up to 458.3 km. We compare our  protocol with the WCP-Ph-QSS and DPS-TF-QSS protocols, which can also surpass the linear rate-distance bound. The results show that our protocol exhibits stronger resistance against the beam splitting attack than both of those protocols. With $\mu=0.84$, the WCP-Ph-QSS protocol has a slightly higher key rate than our protocol, but with  $\mu=1.5$, our protocol has a key rate 5.4  times of that in the WCP-Ph-QSS protocol. Its maximal communication distance (441.7 km) is about 7.9\% longer than that of the WCP-Ph-QSS protocol. DPS-TF-QSS cannot obtain positive key rate with $\mu>0.5074$. It indicates that our protocol exhibits stronger advantages than WCP-Ph-QSS and DPS-TF-QSS protocols in high $\mu$ scenario.
Our dual-DOF QSS protocol offers a promising approach for long-distance and high-capacity quantum networks.

\textbf{Supplementary Material}\\
The key leakage rate, key generation process in ideal and nonideal scenarios, and the parameter estimation.

\begin{acknowledgments}
This work was supported by the National Natural Science Foundation of China under Grant Nos. 92365110 and 12574393.
\end{acknowledgments}

\textbf{Author Declarations}\\
\textbf{Conflict of Interest}\\
The authors have no conflicts to disclose.

\textbf{Data Availability}
The data that support the findings of this study are available
within the article (and its supplementary material).


\begin{thebibliography}{99}
\bibitem{qkd1}
C. H. Bennett and G. Brassard, ``Quantum cryptography: Public key distribution and coin tossing,'' \href{https://doi.org/10.1016/j.tcs.2014.05.025}{Theor. Comput. Sci. \textbf{560}, 7-11 (2014)}.

\bibitem{qkd2}
 A. K. Ekert, ``Quantum cryptography based on Bell's theorem,'' \href{ https://doi.org/10.1103/PhysRevLett.67.661}{Phys. Rev. Lett. \textbf{67}, 661 (1991)}.

\bibitem{qkd3}
 F. Xu, X. Ma,  Q. Zhang,  H. K. Lo and J. W. Pan, ``Secure quantum key distribution with realistic devices,'' \href{ https://doi.org/10.1103/RevModPhys.92.025002}{Rev. Mod. Phys. \textbf{92}, 025002 (2020)}.

\bibitem{qkd5}
 H. K. Lo, M. Curty and B. Qi, ``Measurement-device-independent quantum key distribution,'' \href{https://doi.org/10.1103/PhysRevLett.108.130503}{Phys. Rev. Lett. \textbf{108}, 130503 (2012)}.

 \bibitem{qkd6} Y. M. Bian, Y. Pan, X. S. Xu, L. Zhao, Y. Li, W. Huang, L. Zhang, S. Yu, Y. C. Zhang, and B. J. Xu, ``Continuous-variable quantum key distribution over
  28.6 km fiber with an integrated silicon photonic receiver chip,'' \href{https://doi.org/10.1063/5.0203130}{Appl. Phys. Lett. \textbf{124}, 174001 (2024)}.

\bibitem{qkd7} Y. Pelet, G. Sauder, S. Tanzilli, O. Alibart, and A. Martin,  ``Entanglement-based clock syntonization for quantum key distribution networks: Demonstration over a 50km-long link,'' \href{https://doi.org/10.1063/5.0256758}{Appl. Phys. Lett. \textbf{126}, 174003 (2025)}.

\bibitem{qkd8} A. Ponosova, I. Zhluktova, D. Ruzhitskaya, D. Trefilov, A. Q. Huang, A. Wolf, V. Kamynin, V. Tsvetkov, and V. Makarov, ``Pulsed laser attack at 1061 nm
potentially compromises quantum key distribution,'' \href{https://doi.org/10.1063/5.0287448}{Appl. Phys. Lett. \textbf{127}, 194002 (2025)}.
\bibitem{qss_first} M. Hillery, V. Buzek and A. Berthiaume, ``Quantum secret sharing,'' \href{https://doi.org/10.1103/PhysRevA.59.1829}{Phys. Rev. A \textbf{59}, 1829 (1999)}.

\bibitem{qss1}
A. Karlsson, M. Koashi and N. Imoto, ``Quantum entanglement for secret sharing and secret splitting,'' \href{ https://doi.org/10.1103/PhysRevA.59.162}{Phys. Rev. A \textbf{59}, 162 (1999)}.

\bibitem{qss2}
R. Cleve, D. Gottesman and H. K. Lo, ``How to share a quantum secret,'' \href{ https://doi.org/10.1103/PhysRevLett.83.648}{Phys. Rev. Lett. \textbf{83}, 648 (1999)}.

\bibitem{qss3}
G. P. Guo and G. C. Guo, ``Quantum secret sharing without entanglement,'' \href{ https://doi.org/10.1016/S0375-9601(03)00074-4}{Phys. Rev. A \textbf{310}, 247-251 (2003)}.

\bibitem{qsdc1} G. L. Long, and X. S. Liu, ``Theoretically efficient high-capacity quantum-key-distribution scheme,''
 \href{https://doi.org/10.1103/PhysRevA.65.032302}{Phys. Rev. A} \textbf{65}, 032302 (2002).

\bibitem{qsdc2} F. G. Deng, G. L. Long, and X. S. Liu, ``Two-step quantum direct communication protocol using the Einstein-Podolsky-Rosen pair block,'' \href{https://doi.org/10.1103/PhysRevA.68.042317}{Phys. Rev. A} \textbf{68}, 042317 (2003).

\bibitem{qsdc3} C. W. Ding, W. Y. Wang, W. D. Zhang, L. Zhou, and Y. B. Sheng, ``Quantum secure direct communication based on quantum error correction code,'' \href{ https://doi.org/10.1063/5.0245163}{Appl. Phys. Lett. \textbf{126}, 024002 (2025)}.

\bibitem{qsdc4} H. Zeng, M. M. Du, W. Zhong, L. Zhou, and Y. B. Sheng, ``High-capacity device-independent quantum secure direct communication based on hyper-encoding,'' \href{https://doi.org/10.1016/j.fmre.2023.11.006}{Funda. Res.} \textbf{4}, 852 (2024).

\bibitem{qss5}
Z. J. Zhang and Z. X. Man, ``Multiparty quantum secret sharing of classical messages based on entanglement swapping,'' \href{ https://doi.org/10.1103/PhysRevA.72.022303}{Phys. Rev. A \textbf{72}, 022303 (2005)}.


\bibitem{qss6}
D. Markham and B. C. Sanders, ``Graph states for quantum secret sharing,'' \href{https://doi.org/10.1103/PhysRevA.78.042309}{Phys. Rev. A \textbf{78}, 042309 (2008)}.

\bibitem{qss6n}
Y. Fu, H. L. Yin and T. Y. Chen, ``Long-distance measurement-device-independent multiparty quantum communication,'' \href{https://doi.org/10.1103/PhysRevLett.114.090501}{Phys. Rev. Lett. \textbf{114}, 090501 (2015)}.

\bibitem{qss7}
A. Tavakoli, I. Herbauts, M. \.{Z}ukowski and M. Bourennane, ``Secret sharing with a single d-level quantum system,'' \href{https://doi.org/10.1103/PhysRevA.92.030302}{Phys. Rev. A \textbf{92}, 030302 (2015)}.

\bibitem{qss9}
W. P. Grice and B. Qi, ``Quantum secret sharing using weak coherent states,'' \href{https://doi.org/10.1103/PhysRevA.100.022339}{Phys. Rev. A \textbf{100}, 022339 (2019)}.

\bibitem{diqss}
S. Roy and S. Mukhopadhyay, ``Device-independent quantum secret sharing in arbitrary even dimensions,'' \href{https://doi.org/10.1103/PhysRevA.100.012319}{Phys. Rev. A \textbf{100}, 012319 (2019)}.

\bibitem{qss10}
B. P. Williams, J. M. Lukens, N. A. Peters, B. Qi and W. P. Grice, ``Quantum secret sharing with polarization-entangled photon pairs,'' \href{https://doi.org/10.1103/PhysRevA.99.062311}{Phys. Rev. A \textbf{99}, 062311 (2019)}.

\bibitem{qss11}
X. D. Wu, Y. J. Wang and D. Huang, ``Passive continuous-variable quantum secret sharing using a thermal source,'' \href{https://doi.org/10.1103/PhysRevA.101.022301}{Phys. Rev. A \textbf{101} 022301 (2020)}.

\bibitem{qss12}
Y. Ouyang, K. Goswami, J. Romero, B. C. Sanders, M. H. Hsieh and M. Tomamichel, ``Approximate reconstructability of quantum states and noisy quantum secret sharing schemes,'' \href{https://doi.org/10.1103/PhysRevA.108.012425}{Phys. Rev. A \textbf{108}, 012425 (2023)}.

\bibitem{diqss1}
Q. Zhang, W. Zhong, M. M. Du, S. T. Shen, X. Y. Li, A. L. Zhang, L. Zhou and Y. B. Sheng, ``Device-independent quantum secret sharing with noise preprocessing and postselection,'' \href{https://doi.org/10.1103/PhysRevA.110.042403}{Phys. Rev. A \textbf{110},  042403 (2024)}.

\bibitem{diqss2}
Q. Zhang, J. -W. Ying, Z. J. Wang, W. Zhong, M. M. Du, S. T. Shen, X. Y. Li, A. L. Zhang, S. P. Gu, X. F. Wang, \emph{et al.}, ``Device-independent quantum secret sharing with advanced random key generation basis,'' \href{https://doi.org/10.1103/PhysRevA.111.012603}{Phys. Rev. A \textbf{111}, 012603 (2025)}.

\bibitem{mdiqss2}
C. Zhang, Q. Zhang, W. Zhong, M. M. Du, S. T. Shen, X. Y. Li, A. L. Zhang, L. Zhou and Y. B. Sheng, ``Memory-assisted measurement-device-independent quantum secret sharing,'' \href{https://doi.org/10.1103/PhysRevA.111.012602}{Phys. Rev. A \textbf{111}, 012602 (2025)}.


\bibitem{qss_single_qubit1n}
H. Q. Ma, K. J. Wei and J. H. Yang, ``Experimental single qubit quantum secret sharing in a fiber network configuration,''\href{https://doi.org/10.1364/OL.38.004494}{Opt. Lett. \textbf{38}, 4494-4497 (2013)}.

\bibitem{qss_single_qubit2}
A. Tavakoli, I. Herbauts, M. \.{Z}ukowski and M. Bourennane, ``Secret sharing with a single d-level quantum system,'' \href{https://doi.org/10.1103/PhysRevA.92.030302}{Phys. Rev. A \textbf{92}, 030302 (2015)}.

\bibitem{qss_single_qubit3}
C. Lu, F. Miao, K. Meng and Y. Yu, ``Threshold quantum secret sharing based on single qubit,'' \href{https://doi.org/10.1007/s11128-017-1793-6}{Quantum Information Processing \textbf{17}, 1-13 (2018)}.

\bibitem{qss_experiment1}
Y. A. Chen, A. N. Zhang, Z. Zhao, X. Q. Zhou, C. Y. Lu, C. Z. Peng, T. Yang and J. W. Pan, ``Experimental quantum secret sharing and third-man quantum cryptography," \href{  https://doi.org/10.1103/PhysRevLett.95.200502}{Phys. Rev. Lett. \textbf{95}, 200502 (2005)}.

\bibitem{QSSe3}
C. Schmid, P. Trojek, M. Bourennane, C. Kurtsiefer, M. \.{Z}ukowski and H. Weinfurter, ``Experimental single qubit quantum secret sharing,'' \href{https://doi.org/10.1103/PhysRevLett.95.230505}{Phys. Rev. Lett. \textbf{95}, 230505 (2005)}.

\bibitem{QSSe4}
S. Gaertner, C. Kurtsiefer, M. Bourennane and H. Weinfurter, ``Experimental demonstration of four-party quantum secret sharing,'' \href{https://doi.org/10.1103/PhysRevLett.98.020503}{Phys. Rev. Lett. \textbf{98}, 020503 (2007)}.

\bibitem{qss_experiment4}
B. A. Bell, D. Markham, D.Herrera-Mart\'{\i}, A. Marin, W. Wadsworth, J. Rarity and M.Tame, ``Experimental demonstration of graph-state quantum secret sharing,'' \href{https://doi.org/10.1038/ncomms6480}{Nat. Commun. \textbf{5}, 1-12 (2014)}.

\bibitem{qss_experiment5}
Y. Cai, J. Roslund, G. Ferrini, F. Arzani, X. Xu, C. Fabre and N. Treps, ``Multimode entanglement in reconfigurable graph states using optical frequency combs,'' \href{https://doi.org/10.1038/ncomms15645}{Nat. Commun. \textbf{8}, 15645 (2017)}.

\bibitem{qss_experiment2}
Y. Zhou, J. Yu, Z. Yan, X. Jia, J. Zhang, C. Xie and K. Peng, ``Quantum secret sharing among four players using multipartite bound entanglement of an optical field,'' \href{https://doi.org/10.1103/PhysRevLett.121.150502}{Phys. Rev. Lett. \textbf{121}, 150502 (2018)}.

\bibitem{QSSe8}
Y. R. Xiao, H. L. Yin, W. J. Hua, X. Y. Cao and Z. B. Chen, ``Experimental Efficient Source-Independent Quantum Secret Sharing against Coherent Attacks,'' \href{https://doi.org/10.1103/7wrk-p6gl}{Phys. Rev. Lett. \textbf{135}, 150801 (2025)}.

\bibitem{linear_rate4} S. Pirandola1, R. Garc\'{\i}a-Patr\'{o}n1, S. L. Braunstein, and S. Lloyd, ``Direct and reverse secret-key capacities of a quantum channel,'' \href{ https://doi.org/10.1103/PhysRevLett.102.050503}{Phys. Rev. Lett \textbf{102}, 050503 (2009)}.

\bibitem{linear_rate1}
M. Takeoka, S. Guha and M. M. Wilde, ``Fundamental rate-loss tradeoff for optical quantum key distribution,'' \href{https://doi.org/10.1038/ncomms6235}{Nat. Commun. \textbf{5}, 5235 (2014)}.

\bibitem{linear_rate3}
S. Pirandola, R. Laurenza, C. Ottaviani and L. Banchi, ``Fundamental limits of repeaterless quantum communications,'' \href{https://doi.org/10.1038/ncomms15043}{Nat. Commun. \textbf{8}, 15043 (2017)}.

\bibitem{linear_rate3n} R. Laurenza, S. Pirandola, ``General bounds for sender-receiver capacities in multipoint quantum communications,''
  \href{ https://doi.org/10.1103/PhysRevA.96.032318}{Phys. Rev. A \textbf{96}, 032318 (2017)}.  

\bibitem{linear_rate2n} S. Pirandola, ``General upper bounds for distributing conferencing keys in arbitrary quantum networks,''  \href{https://doi.org/10.1049/iet-qtc.2020.0006}{IET Quantum Commun. \textbf{1}, 22-25 (2020)}.



\bibitem{TF-QKD3}
X. B. Wang, Z. W. Yu and X. L. Hu, ``Twin-field quantum key distribution with large misalignment error,'' \href{https://doi.org/10.1103/PhysRevA.98.062323}{Phys. Rev. A \textbf{98}, 062323 (2018)}.


\bibitem{TF-QKD2}
C. Cui, Z. Q. Yin, R. Wang, W. Chen, S. Wang, G. C. Guo and Z. F. Han, ``Twin-field quantum key distribution without phase postselection,'' \href{https://doi.org/10.1103/PhysRevApplied.11.034053}{Phys. Rev. Appl. \textbf{11}, 034053 (2019)}.

\bibitem{TF-QKD1}
S. Wang, Z. Q. Yin, D. Y. He, W. Chen, R. Q. Wang, P. Ye, Y. Zhou, G. Fan-Yuan, F. X. Wang and W. Chen, \emph{et al}., ``Twin-field quantum key distribution over 830-km fibre,'' \href{https://doi.org/10.1038/s41566-021-00928-2}{Nat. Photonics \textbf{16}, 154-161 (2022)}.

\bibitem{TF-QKD4}
Y. Liu, W. J. Zhang, C. Jiang, J. P. Chen, C. Zhang, W. X. Pan, D. Ma, H. Dong and J. M. Xiong, ``Experimental twin-field quantum key distribution over 1000 km fiber distance,'' \href{https://doi.org/10.1103/PhysRevLett.130.210801}{Phys. Rev. Lett. \textbf{130}, 210801  (2023)}.

\bibitem{DPS_TF_QSS}
J. Gu, X. Y. Cao, H. L. Yin and Z. B. Chen, ``Differential phase shift quantum secret sharing using a twin field,'' \href{https://doi.org/10.1364/OE.417856}{Opt. Express \textbf{29}, 9165-9173 (2021)}.

\bibitem{RR_QSS}
J. Gu, Y. M. Xie, W. B. Liu, Y. Fu, H. L. Yin and Z. B. Chen, ``Secure quantum secret sharing without signal disturbance monitoring,'' \href{https://doi.org/10.1364/OE.440365}{Opt. Express \textbf{29}, 32244-32255 (2021)}.

\bibitem{WCP_Ph_QSS}
A. Shen, X. Y. Cao, Y. Wang, Y. Fu, J. Gu, W. B. Liu, C. X. Weng, H. L. Yin and Z. B. Chen, ``Experimental quantum secret sharing based on phase encoding of coherent states,'' \href{https://doi.org/10.1007/s11433-023-2105-7}{Sci. China Phys. Mech. Astron. \textbf{66}, 260311 (2023)}.

\bibitem{high-dimension1}
J. Pinnell, I. Nape, M. de Oliveira, N. TabeBordbar and A. Forbes, ``Experimental demonstration of 11-dimensional 10-party quantum secret sharing,'' \href{ https://doi.org/10.1002/lpor.202000012}{Laser Photonics Rev. \textbf{14}, 2000012 (2020)}.

\bibitem{high-dimension}
M. De Oliveira, I. Nape, J. Pinnell, N. TabeBordbar and A. Forbes, ``Experimental high-dimensional quantum secret sharing with spin-orbit-structured photons,'' \href{https://doi.org/10.1103/PhysRevA.101.042303}{Phys. Rev. A \textbf{101}, 042303 (2020)}.

\bibitem{INI_qkd}
Y. Yu, W. Li, L. Wang and S. M. Zhao, ``Interfering-or-not-interfering quantum key distribution,'' \href{https://doi.org/10.1103/PhysRevA.109.052609}{Phys. Rev. A \textbf{109}, 052609 (2024)}.

\bibitem{BSA1}
J. Calsamiglia, S. M. Barnett and N. L\"{u}tkenhaus, ``Conditional beam-splitting attack on quantum key distribution,'' \href{https://doi.org/10.1103/PhysRevA.65.012312}{Phys. Rev. A \textbf{65}, 012312 (2001)}.

\bibitem{BSA2}
X. Ma, P. Zeng and H. Zhou, ``Phase-matching quantum key distribution,'' \href{https://doi.org/10.1103/PhysRevX.8.031043}{Phys. Rev. X \textbf{8}, 031043 (2018)}.

\end{thebibliography}
\end{document}


\title{Supplementary Material\\High capacity dual degrees of freedom quantum secret sharing protocol beyond the linear rate-distance bound}
\author{
	Meng-Dong Zhu$^{1,2}$, 
	Cheng Zhang$^{2}$, 
	Shi-Pu Gu$^{2}$, 
	Xing-Fu Wang$^{1}$, 
	Lan Zhou$^{1,3}$\footnote{Email address: zhoul@njupt.edu.cn}, 
	and Yu-Bo Sheng$^{2,3}$\footnote{Email address: shengyb@njupt.edu.cn}
}
\address{
	$^1$College of Science, Nanjing University of Posts and Telecommunications, Nanjing, Jiangsu 210023, China\\
	$^2$College of Electronic and Optical Engineering and College of Flexible Electronics (Future Technology), Nanjing University of Posts and Telecommunications, Nanjing, Jiangsu 210023, China\\
	$^3$School of Physics, Hangzhou Normal University, Hangzhou, Zhejiang 311121, China
}
\date{\today}
\maketitle

\section{The key leakage of the dual-DOF QSS protocol under the beam-splitting attack}
 Under the beam splitting attack, the eavesdropper (Eve) uses a variable beam splitter (VBS) on the channel from Alice to Charlie. The reflected pulses are stored in his quantum memory (QM) and the transmitted pulses are sent to Charlie through the ideal lossless quantum channels. As the transmittance of the VBS precisely matches the transmittance efficiency $\eta_t$ of the actual quantum channel and the eavesdropping does not change the encoded bits of the pulses, this attack will not influence the gains in Charlie's photon detectors or increase the QBER of the security checking. As all of Bob's key bits should be announced, Eve does not require to attack Bob's quantum channel. After Alice announcing the generation basis for each WCP and the locations of the key bits encoded WCPs, Eve extracts the corresponding photons to make the unambiguous state discrimination (USD) measurement \cite{init_USD} to read out Alice's key bits.

USD is a quantum measurement technology designed to distinguish a set of nonorthogonal quantum states.
USD cannot succeed with the probability of 100\% and its maximum success probability is inherently bounded by the overlap between the states. For the simplest case of two nonorthogonal states, $|S_0\rangle$ and $|S_1\rangle$, the maximum success probability of the USD measurement is given by $1-|\langle S_0|S_1 \rangle|$. In a more general case for a set of nonorthogonal states, the USD measurement's success probability ($P_{\mathrm{success}}$)  can be upper bounded by \cite{PM_QKD,USD}.
\begin{equation}
P_{\mathrm{success}}\leqslant1-\frac{1}{N-1}\sum_{i\neq j}\sqrt{p_ip_j}|\langle\psi_i\mid\psi_j\rangle|,
\end{equation}
where $N$ represents the number of nonorthogonal quantum states, and $p_{i(j)}$ corresponds to the preparation probability of the state $|\psi_{i(j)}\rangle$.

In our protocol, the four states prepared by Alice or Bob in the X basis are
\begin{eqnarray}
|\psi_0\rangle&=&|\sqrt{(1-\eta_t)\mu}\rangle_+
=|\sqrt{(1-\eta_t)\mu/2}\rangle_H
+|\sqrt{(1-\eta_t)\mu/2}\rangle_V,\nonumber\\
|\psi_1\rangle&=&|-\sqrt{(1-\eta_t)\mu}\rangle_+
=|-\sqrt{(1-\eta_t)\mu/2}\rangle_H
+|-\sqrt{(1-\eta_t)\mu/2}\rangle_V,\nonumber\\
|\psi_2\rangle&=&|\sqrt{(1-\eta_t)\mu}\rangle_-
=|\sqrt{(1-\eta_t)\mu/2}\rangle_H
+|-\sqrt{(1-\eta_t)\mu/2}\rangle_V,\nonumber\\
|\psi_3\rangle&=&|-\sqrt{(1-\eta_t)\mu}\rangle_-
=|-\sqrt{(1-\eta_t)\mu/2}\rangle_H
+|\sqrt{(1-\eta_t)\mu/2}\rangle_V,
\end{eqnarray}
where $\mu$ is the average photon number of the WCPs. In our protocol, we can obtain $N=4$ and $p_i=p_j=\frac{1}{4}$. Based on the inner product formula of coherent states
$\langle\alpha|\beta\rangle=e^{-|\alpha|^2/2-|\beta|^2/2+\alpha^*\beta}$, $P_{\mathrm{success}}$ can be upper bounded by
\begin{equation}
P_{\mathrm{success}}\leq1-\frac{1}{3}[e^{-2(1-\eta_{t})\mu}+2e^{-(1-\eta_{t})\mu}].
\end{equation}

In our protocol, Eve can stolen the encoded key bits as long as the USD measurements is successful,
so that the key leakage rate ($I_E$) to Eve is
\begin{equation}
I_E=P_{\mathrm{success~}}\leq1-\frac{1}{3}[e^{-2(1-\eta_{t})\mu}+2e^{-(1-\eta_{t})\mu}].
\end{equation}

Meanwhile, we take some other examples for comparison, such as DPS-TF-QSS \cite{DPS_TF_QSS}, WCP-Ph-QSS \cite{WCP_Ph_QSS} and MDI-QKD (WCP-Pol-QSS) \cite{qkd5}.
Under the beam splitting attack, $I_E$ of the WCP-Ph-QSS and WCP-Pol-QSS protocols can be upper bounded by \cite{PM_QKD,qkd5}
\begin{eqnarray}
I^{WCP-Ph-QSS}_E&\leq&1-e^{-2(1-\eta_t)\mu},\nonumber\\
I^{WCP-Pol-QSS}_E&\leq&1-e^{-(1-\eta_t)\mu}.
\end{eqnarray}
For the DPS-TF-QSS protocol, instead of using the USD measurement, collision probability analysis is adopted because its bit correlation depends on the time slots. Even if the USD measurement is applied in this protocol, its key leakage to Eve is still within the  range calculated by collision probability.
As a result, we can obtain $I_E$ of the DPS-TF-QSS as \cite{DPS_TF_QSS,Secruity_DPS_TF_QSS}
 \begin{equation}
 I^{DPS-TF-QSS}_E\leq2\mu(1 - \eta_t).
\end{equation}

\section{Key generation process and phase error rate of our dual-DOF QSS protocol in ideal scenario}\label{sec222}
Suppose that Alice's WCP in $X_{dual}$ basis has the quantum state as
\begin{align}
|\psi_{A}\rangle=\frac{1}{2}\Bigl(
 |0\rangle_{a}^{Pol}|0\rangle_{a}^{Ph}|\sqrt{\mu}\rangle_{A_+} + |0\rangle_{a}^{Pol}|1\rangle_{a}^{Ph}|-\sqrt{\mu}\rangle_{A_+}
 + |1\rangle_{a}^{Pol}|0\rangle_{a}^{Ph}|\sqrt{\mu}\rangle_{A_-} + |1\rangle_{a}^{Pol}|1\rangle_{a}^{Ph}|-\sqrt{\mu}\rangle_{A_-} \Bigr),
\end{align}
where $|0 (1)\rangle_{a}^{Pol(Ph)}$ represents the key bit 0 (1) in the polarization (phase) DOF, and $|\pm\sqrt{\mu} \rangle_{A_\pm}$ denote the coherent state prepared in $X_{dual}$ basis. Similarly, Bob's WCP in $X_{dual}$ basis can be written as
\begin{align}
|\psi_{B}\rangle=\frac{1}{2}\Bigl(
 |0\rangle_{b}^{Pol}|0\rangle_{b}^{Ph}|\sqrt{\mu}\rangle_{B_+} + |0\rangle_{b}^{Pol}|1\rangle_{b}^{Ph}|-\sqrt{\mu}\rangle_{B_+}
 + |1\rangle_{b}^{Pol}|0\rangle_{b}^{Ph}|\sqrt{\mu}\rangle_{B_-} + |1\rangle_{b}^{Pol}|1\rangle_{b}^{Ph}|-\sqrt{\mu}\rangle_{B_-} \Bigr).
\end{align}

We take the case where both parties encode the bit pair $00$ as an example. We describe the state evolution through the beam splitter (BS) and polarizing beam splitter (PBS)
in Charlie's location without considering channel loss and dark counts as
\begin{align}\label{eq.6}\notag       
&|0\rangle_a^{Pol}|0\rangle_a^{Ph}|\sqrt{\mu}\rangle_{A_+}\otimes|0\rangle_b^{Pol}|0\rangle_b^{Ph}|\sqrt{\mu}\rangle_{B_+}
=\frac{1}            {2}|0\rangle_a^{Pol}|0\rangle_a^{Ph}|0\rangle_b^{Pol}|0\rangle_b^{Ph}(|\sqrt{\mu}\rangle_{A_H}+|\sqrt{\mu}\rangle_{A_V})(|\sqrt{\mu}\rangle_{B_H}+|\sqrt{\mu}\rangle_{B_V})\notag
\\
&\xrightarrow{BS, PBS} \notag
\frac{1}  {4}|0\rangle_a^{Pol}|0\rangle_a^{Ph}|0\rangle_b^{Pol}|0\rangle_b^{Ph}(|\sqrt{\mu}\rangle_{D_{1H}}+|\sqrt{\mu}\rangle_{D_{2H}}+|\sqrt{\mu}\rangle_{D_{1V}}+|\sqrt{\mu}\rangle_{D_{2V}})\\
&\otimes (|\sqrt{\mu}\rangle_{D_{1H}}-|\sqrt{\mu}\rangle_{D_{2H}}+|\sqrt{\mu}\rangle_{D_{1V}}-|\sqrt{\mu}\rangle_{D_{2V}})
=|0\rangle_a^{Pol}|0\rangle_a^{Ph}|0\rangle_b^{Pol}|0\rangle_b^{Ph}|\sqrt{\mu}\rangle_{D_{1H}}|\sqrt{\mu}\rangle_{D_{1V}}.
\end{align}

For simplicity, we rewrite $|0\rangle_a^{Pol}|0\rangle_a^{Ph}|0\rangle_b^{Pol}|0\rangle_b^{Ph}|\sqrt{\mu}\rangle_{D_{1H}}|\sqrt{\mu}\rangle_{D_{1V}}$ as $|00\rangle_a|00\rangle_{b}|\sqrt{\mu}\rangle_{H_1}|\sqrt{\mu}\rangle_{V_1}$.  Here, we present a summary of the detector responses of all 16 possible combinations of Alice's and Bob's states in $X_{dual}$ basis as
\begin{equation}\label{eq.31}
\begin{aligned}[b]
&\frac{1}{4}[|00\rangle_a|00\rangle_{b} |\sqrt{\mu}\rangle_{H_1}|\sqrt{\mu}\rangle_{V_1}
    + |00\rangle_a|01\rangle_{b}  |\sqrt{\mu}\rangle_{H_2}|\sqrt{\mu}\rangle_{V_2}
    +  |00\rangle_a|10\rangle_{b}  |\sqrt{\mu}\rangle_{H_1}|\sqrt{\mu}\rangle_{V_2}\\
    &+ |00\rangle_a|11\rangle_{b}  |\sqrt{\mu}\rangle_{V_1}|\sqrt{\mu}\rangle_{H_2}
    +  |01\rangle_a|00\rangle_{b}  |{-}\sqrt{\mu}\rangle_{H_2}|{-}\sqrt{\mu}\rangle_{V_2}
    + |01\rangle_a|01\rangle_{b}  |{-}\sqrt{\mu}\rangle_{H_1}|{-}\sqrt{\mu}\rangle_{V_1} \\
    &+  |01\rangle_a|10\rangle_{b}  |{-}\sqrt{\mu}\rangle_{V_1}|{-}\sqrt{\mu}\rangle_{H_2}
    + |01\rangle_a|11\rangle_{b}  |{-}\sqrt{\mu}\rangle_{H_1}|{-}\sqrt{\mu}\rangle_{V_2}
    +  |10\rangle_a|00\rangle_{b}  |\sqrt{\mu}\rangle_{H_1}|{-}\sqrt{\mu}\rangle_{V_2}\\
    &+ |10\rangle_a|01\rangle_{b}  |{-}\sqrt{\mu}\rangle_{V_1}|\sqrt{\mu}\rangle_{H_2}
    +  |10\rangle_a|10\rangle_{b}  |\sqrt{\mu}\rangle_{H_1}|{-}\sqrt{\mu}\rangle_{V_1}
    + |10\rangle_a|11\rangle_{b}  |\sqrt{\mu}\rangle_{H_2}|{-}\sqrt{\mu}\rangle_{V_2} \\
    &+  |11\rangle_a|00\rangle_{b}  |\sqrt{\mu}\rangle_{V_1}|{-}\sqrt{\mu}\rangle_{H_2}
    + |11\rangle_a|01\rangle_{b}  |{-}\sqrt{\mu}\rangle_{H_1}|\sqrt{\mu}\rangle_{V_1}
    +  |11\rangle_a|10\rangle_{b}  |{-}\sqrt{\mu}\rangle_{H_2}|\sqrt{\mu}\rangle_{V_2}\\
    &+ |11\rangle_a|11\rangle_{b}  |{-}\sqrt{\mu}\rangle_{H_1}|\sqrt{\mu}\rangle_{V_2}].
\end{aligned}
\end{equation}

We analyze the impact of coherent states on detector responses by dividing the coherent state into a superposition of the vacuum state and non-vacuum states with the form of
\begin{eqnarray}\label{eq32}
|\sqrt\mu\rangle&=&e^{-\frac{\mu}{2}}\sum_{i=0}^{\infty}\sqrt{\frac{\mu^i}{i!}}\left|i\right\rangle
=e^{-\frac{\mu}{2}}\left|0\right\rangle+e^{-\frac{\mu}{2}}\sum_{i=1}^{\infty}\sqrt{\frac{\mu^i}{i!}}\left|i\right\rangle\notag\\
&=&e^{-\frac{\mu}{2}}\left|0\right\rangle+\left|\sqrt{\mu^\prime}\right\rangle.
\end{eqnarray}
 Three types of detector responses can be induced: no response, one detector response and two detectors response. As the no response case cannot be used to generate keys, we only analyze the one detector response and two detectors response cases.


\subsection{The key generation and phase error of the BSM results in Event$_1$}
We first consider the case that $D_{1H}$ responds. Based on Eq. (\ref{eq.31}) and Eq. (\ref{eq32}), the items with only $D_{1H}$ responding  can be written as
\begin{equation}\begin{aligned}\label{e1}
 & e^{\frac{-\mu}{2}}[|00\rangle_a|00\rangle_{b}|\sqrt{\mu^{\prime}}\rangle_{H_1}
 +|00\rangle_a|10\rangle_{b}|\sqrt{\mu^{\prime}}\rangle_{H_1}
 +|01\rangle_a|01\rangle_{b}|-\sqrt{\mu^{\prime}}\rangle_{H_1}
  +|01\rangle_a|11\rangle_{b}|-\sqrt{\mu^{\prime}}\rangle_{H_1}\\
&+|10\rangle_a|00\rangle_{b}|\sqrt{\mu^{\prime}}\rangle_{H_1}
+|10\rangle_a|10\rangle_{b}|\sqrt{\mu^{\prime}}\rangle_{H_1}
+|11\rangle_a|01\rangle_{b}|-\sqrt{\mu^{\prime}}\rangle_{H_1}+|11\rangle_{a}|11\rangle_{b}|-\sqrt{\mu^{\prime}}\rangle_{H_1}].
\end{aligned}\end{equation}

We can rewrite the quantum state in Eq. (\ref{e1}) as
\begin{equation}\begin{aligned}
 & \frac{1}{\sqrt{2}}\left(|00\rangle_{ab}^{\mathrm{Pol}}+|11\rangle_{ab}^{\mathrm{Pol}}\right)\otimes\frac{1}{\sqrt{2}}\left[|00\rangle_{ab}^{\mathrm{Ph}}+(-1)^m|11\rangle_{ab}^{\mathrm{Ph}}\right] +\frac{1}{\sqrt{2}}(|01\rangle_{ab}^{\mathrm{Pol}}+|10\rangle_{ab}^{\mathrm{Pol}})\otimes\frac{1}{\sqrt{2}}\left[|00\rangle_{ab}^{\mathrm{Ph}}+(-1)^m|11\rangle_{ab}^{\mathrm{Ph}}\right] \\
 &=\frac{1}{\sqrt{2}}\left(\left|\Phi^+\right\rangle_{ab}^{\mathrm{Pol}}+\left|\Psi^+\right\rangle_{ab}^{\mathrm{Pol}}\right)\otimes\frac{1}{\sqrt{2}}\left[|00\rangle_{ab}^{\mathrm{Ph}}+(-1)^m|11\rangle_{ab}^{\mathrm{Ph}}\right] ,
\end{aligned}\end{equation}
where $m$ denotes the photon number in the coherent state that arrives at $D_{1H}$ ($m>0$). On one hand, it can be observed that only the bit's correlation in the phase DOF can be deduced from the detector response. Charlie defines the key bit in the phase DOF as 0. Charlie cannot deduce the bit's correlation in the polarization DOF from the detector response.

On the other hand, it can be found that the Bell state in the phase DOF is uncertain. When $m$ is even, the Bell state is $|\Phi^+\rangle_{ab}^{\mathrm{Ph}}$, while when $m$ is odd, the Bell state is $|\Phi^-\rangle_{ab}^{\mathrm{Ph}}$. This uncertainty provides Eve an opportunity for key stealing.
Considering the poisson distribution characteristics of the coherent state, the sum of the probabilities for odd photon number items among integers ($m > 0$) exceeds the sum of those for the even photon number items. For reducing the phase error rate, we define that the BSM results corresponding to odd number (o) of $m$ are the correct BSM results and those corresponding to even number (e) of $m$ have a phase error. The details of the correct BSM results and the BSM results corresponding to phase errors under different $m$ in Event$_1$ are shown in Tab. \ref{table2}.
\begin{table}[htbp]
\centering
\caption{The correct BSM results and the wrong BSM results corresponding to the phase
error under $m$ in Event$_1$. In this case, the number of phase error is 1. }
\setlength{\tabcolsep}{6pt}
\begin{tabular}{ccc}
\doubletoprule
\addlinespace[1.5ex]
\multicolumn{1}{c}{Detector response}&\multicolumn{1}{c}{$m$} & \multicolumn{1}{c}{BSM results} \\
\midrule
 $D_{1H}$ & odd  & \makecell{$\ket{\Phi^{-}}_{ab}^{Ph}$ \\
correct BSM result} \\
 \addlinespace[1.5ex]
 $D_{1H}$ & even & \makecell{$\ket{\Phi^{+}}_{ab}^{Ph}$ \\
phase error:1} \\
 \addlinespace[1.5ex]
  $D_{2H}$ & odd & \makecell{$\ket{\Psi^{-}}_{ab}^{Ph}$ \\
correct BSM result} \\
 \addlinespace[1.5ex]
  $D_{2H}$ & even & \makecell{$\ket{\Psi^{+}}_{ab}^{Ph}$ \\
phase error:1} \\
 \addlinespace[1.5ex]
\doublebottomrule
\end{tabular}
\label{table2}
\end{table}

When only $D_{2H}$ responds, the corresponding items  can be written as
\begin{equation}\begin{aligned}
 & e^{\frac{-\mu}{2}}[|00\rangle_a|01\rangle_{b}|\sqrt{\mu^{\prime}}\rangle_{H_2}
+|00\rangle_a|11\rangle_{b}|\sqrt{\mu^{\prime}}\rangle_{H_2}
 +|01\rangle_a|00\rangle_{b}|-\sqrt{\mu^{\prime}}\rangle_{H_2}
  +|01\rangle_a|10\rangle_{b}|-\sqrt{\mu^{\prime}}\rangle_{H_2}\\
&+|10\rangle_a|01\rangle_{b}|\sqrt{\mu^{\prime}}\rangle_{H_2}
+|10\rangle_a|11\rangle_{b}|\sqrt{\mu^{\prime}}\rangle_{H_2}
+|11\rangle_a|00\rangle_{b}|-\sqrt{\mu^{\prime}}\rangle_{H_2}+|11\rangle_a|10\rangle_{b}|-\sqrt{\mu^{\prime}}\rangle_{H_2}],
\end{aligned}\end{equation}
and the bit's correlation  can be represented as
\begin{equation}
    \frac{1}{\sqrt{2}}\left(\left|\Phi^+\right\rangle_{ab}^{\mathrm{Pol}}+\left|\Psi^+\right\rangle_{ab}^{\mathrm{Pol}}\right)\otimes\frac{1}{\sqrt{2}}\left[|01\rangle_{ab}^{\mathrm{Ph}}+(-1)^m|10\rangle_{ab}^{\mathrm{Ph}}\right].
\end{equation}

Similarly, Charlie obtains the key bit in the phase DOF as 1. We also define that the BSM results corresponding to odd number (o) of $m$ are the correct BSM results and those corresponding to even number (e) of $m$ have a phase error. In the following calculations, the phase error probability corresponding to $D_{1H}$ responding is defined as $P_r^{H_1}(e)$, and that corresponding to $D_{2H}$ responding is defined as $P_r^{H_2}(e)$.

\subsection{The key generation and phase error of the BSM results in Event$_2$}
Based on Eq. (\ref{eq.31}) and Eq. (\ref{eq32}), the items with $D_{1H}$ and $D_{1V}$ responding can be written as
\begin{equation}
\begin{aligned}
&|00\rangle_a|00\rangle_{b}|\sqrt{\mu^\prime}\rangle_{H_1}|\sqrt{\mu^\prime}\rangle_{V_1}
+|01\rangle_a|01\rangle_{b}|-\sqrt{\mu^\prime}\rangle_{H_1}|-\sqrt{\mu^\prime}\rangle_{V_1}\\
&+|10\rangle_a|10\rangle_{b}|\sqrt{\mu^\prime}\rangle_{H_1}|-\sqrt{\mu^\prime}\rangle_{V_1}
+|11\rangle_a|11\rangle_{b}|-\sqrt{\mu^\prime}\rangle_{H_1}|\sqrt{\mu^\prime}\rangle_{V_1},
\end{aligned}\label{e2}
\end{equation}
which can be rewritten as
\begin{equation}
\begin{aligned}
\left|00\right\rangle_{ab}^{Pol}{[\left|00\right\rangle}_{ab}^{Ph}+{(-1)}^{(m+n)}\left|11\right\rangle_{ab}^{Ph}]
+{\left|11\right\rangle_{ab}^{Pol}[(-1)}^n\left|00\right\rangle_{ab}^{Ph}+{(-1)}^m\left|11\right\rangle_{ab}^{Ph}].
\end{aligned}
\end{equation}
 Here, $m$ and $n$ denote the photon number of the coherent states arriving at $D_{1H}$ and $D_{1V}$, respectively. It can be found that Alice's and Bob's key bits in polarization and phase DOFs are both identical, so that Charlie can obtain the key bits as 00. Meanwhile, similar as the definition in Event$_1$,
we define the correct BSM results corresponding to both $m$, $n$ being odd numbers, while the BSM results corresponding to other different combinations of $m$ and $n$  are defined as phase error. The details of the correct BSM results and the BSM results corresponding to phase errors under different $m$ and $n$ in Event$_2$ are shown in Tab. \ref{table3}.
\begin{table}[htbp]
\centering
\caption{The correct BSM results and BSM results corresponding to phase
error under different even-odd combinations of $m$ and $n$ in Event$_2$.  The number of phase error may be 1 or 2.}
\setlength{\tabcolsep}{6pt}
\begin{tabular}{cccc} 
\doubletoprule
\addlinespace[1.5ex]
\multirow{2}{*}{\makecell{Detector\\responses}} & \multirow{2}{*}{$n (D_{1(2)V})$} & \multicolumn{2}{c}{$m (D_{1(2)H})$} \\
\cmidrule(lr){3-4}
& & Odd & Even \\
\midrule
\multirow{1}{*}{$D_{1H}, D_{1V}$} & \multirow{1}{*}{Odd}
& \makecell{$\ket{\Phi^{-}}_{ab}^{Pol}  \ket{\Phi^{+}}_{ab}^{Ph}$ \\
correct BSM result}
& \makecell{$\ket{\Phi^{-}}_{ab}^{Pol} \ket{\Phi^{-}}_{ab}^{Ph}$ \\
phase error: 1} \\
\addlinespace[1.5ex]
\multirow{1}{*}{$D_{1H}, D_{1V}$} & \multirow{1}{*}{Even}
& \makecell{$\ket{\Phi^{+}}_{ab}^{Pol}
\ket{\Phi^{-}}_{ab}^{Ph}$ \\
phase error: 2}
& \makecell{$\ket{\Phi^{+}}_{ab}^{Pol}  \ket{\Phi^{+}}_{ab}^{Ph}$ \\
phase error: 1} \\
\addlinespace[1.5ex]
\multirow{1}{*}{$D_{2H}, D_{2V}$} & \multirow{1}{*}{Odd}
& \makecell{$\ket{\Phi^{-}}_{ab}^{Pol}  \ket{\Psi^{+}}_{ab}^{Ph}$ \\
correct BSM result}
& \makecell{$\ket{\Phi^{-}}_{ab}^{Ph}  \ket{\Psi^{-}}_{ab}^{Ph}$\\
phase error: 1} \\
\addlinespace[1.5ex]
\multirow{1}{*}{$D_{2H}, D_{2V}$} & \multirow{1}{*}{Even}
& \makecell{$\ket{\Phi^{+}}_{ab}^{Pol}
\ket{\Psi^{-}}_{ab}^{Ph}$ \\
phase error: 2}
& \makecell{$\ket{\Phi^{+}}_{ab}^{Pol}  \ket{\Psi^{+}}_{ab}^{Ph}$ \\
phase error: 1} \\
\addlinespace[1ex]
\doublebottomrule
\end{tabular}
\label{table3}
\end{table}

Similarly, the items corresponding to $D_{2H}$ and $D_{2V}$ responding can be written as
\begin{equation}
\begin{aligned}
&|00\rangle_a|01\rangle_{b}|\sqrt{\mu^\prime}\rangle_{H_2}|\sqrt{\mu^\prime}\rangle_{V_2}
+|01\rangle_a|00\rangle_{b}|-\sqrt{\mu^\prime}\rangle_{H_2}|-\sqrt{\mu^\prime}\rangle_{V_2}\\
&+|10\rangle_a|11\rangle_{b}|\sqrt{\mu^\prime}\rangle_{H_2}|-\sqrt{\mu^\prime}\rangle_{V_2}
+|11\rangle_a|10\rangle_{b}|-\sqrt{\mu^\prime}\rangle_{H_2}|\sqrt{\mu^\prime}\rangle_{V_2},
\end{aligned}
\end{equation}
which can be rewritten as
\begin{equation}
\begin{aligned}
\left|00\right\rangle_{ab}^{Pol}{[\left|01\right\rangle}_{ab}^{Ph}+{(-1)}^{(m+n)}\left|10\right\rangle_{ab}^{Ph}]
+{\left|11\right\rangle_{ab}^{Pol}[(-1)}^n\left|01\right\rangle_{ab}^{Ph}+{(-1)}^m\left|10\right\rangle_{ab}^{Ph}].
\end{aligned}
\end{equation}
where $m$ and $n$ denote the photon number of the coherent states arriving at $D_{2H}$ and $D_{2V}$, respectively. It can be found that Charlie can obtain the key bits as 01. We also define the correct BSM results corresponding to both $m$, $n$ being odd numbers.

When detectors $D_{1H}$ and $D_{1V}$ ($D_{2H}$ and $D_{2V}$) respond, the phase error probability corresponding to even numbers of $m$ and $n$ are defined as $P_r^{H_1V_1(H_2V_2)}(e,e)$, that corresponding to even number of $m$ and odd number of $n$ is $P_r^{H_1V_1(H_2V_2)}(e,o)$. Similarly, we can also define the phase error probabilities $P_r^{H_1V_1(H_2V_2)}(o,e)$.

\subsection{The key generation and phase error of the BSM results in Event$_3$}
Based on Eq. (\ref{eq.31}) and Eq. (\ref{eq32}), the items corresponding to $D_{1H}$ and $D_{2V}$ responding have the form of
\begin{equation}
\begin{aligned}
&|00\rangle_a|10\rangle_{b}|\sqrt{\mu^\prime}\rangle_{H_1}|\sqrt{\mu^\prime}\rangle_{V_2}
+|01\rangle_a|11\rangle_{b}|-\sqrt{\mu^\prime}\rangle_{H_1}|-\sqrt{\mu^\prime}\rangle_{V_2}\\
&+|10\rangle_a|00\rangle_{b}|\sqrt{\mu^\prime}\rangle_{H_1}|-\sqrt{\mu^\prime}\rangle_{V_2}
+|11\rangle_a|01\rangle_{b}|-\sqrt{\mu^\prime}\rangle_{H_1}|\sqrt{\mu^\prime}\rangle_{V_2},
\end{aligned}
\end{equation}
which can be rewritten as
\begin{equation}
\begin{aligned}
&\left|01\right\rangle_{ab}^{Pol}{[\left|00\right\rangle}_{ab}^{Ph}+{(-1)}^{(m+n)}\left|11\right\rangle_{ab}^{Ph}]
+{\left|10\right\rangle_{ab}^{Pol}[(-1)}^n\left|00\right\rangle_{ab}^{Ph})+{(-1)}^m\left|11\right\rangle_{ab}^{Ph}].
\end{aligned}\label{e32}
\end{equation}
Here, $m$ and $n$ denote the numbers of photons arriving at $D_{1H}$ and $D_{2V}$, respectively. From Eq. (\ref{e32}), Alice's and Bob's key bits in polarization DOF are different while those in phase DOF are identical, so that Charlie can obtain the key bits as 10.

In the same way, we analyze the phase errors in the BSM results with different odd-even combinations of $m$ and $n$. The details of the correct BSM results and the BSM results corresponding to phase errors under different $m$ and $n$ in Event$_3$ are shown in Tab. \ref{table4}.
\begin{table}[htbp]
\centering
\caption{The correct BSM results and BSM results corresponding to phase
error under different even-odd combinations of $m$ and $n$ in Event$_3$. The number of phase error may be 1 or 2.}
\setlength{\tabcolsep}{6pt}
\begin{tabular}{cccc} 
\doubletoprule
\addlinespace[1.5ex]
\multirow{2}{*}{\makecell{Detector\\responses}} & \multirow{2}{*}{$n (D_{1(2)V})$} & \multicolumn{2}{c}{$m (D_{1(2)H})$} \\
\cmidrule(lr){3-4}
& & Odd & Even \\
\midrule
\multirow{1}{*}{$D_{1H}, D_{2V}$} & \multirow{1}{*}{Odd}
& \makecell{$\ket{\Psi^{-}}_{ab}^{Pol}  \ket{\Phi^{+}}_{ab}^{Ph}$ \\
correct BSM result}
& \makecell{$\ket{\Psi^{-}}_{ab}^{Pol} \ket{\Phi^{-}}_{ab}^{Ph}$ \\
phase error: 1} \\
\addlinespace[1.5ex]
\multirow{1}{*}{$D_{1H}, D_{2V}$} & \multirow{1}{*}{Even}
& \makecell{$\ket{\Psi^{+}}_{ab}^{Pol}
\ket{\Phi^{-}}_{ab}^{Ph}$ \\
phase error: 2}
& \makecell{$\ket{\Psi^{+}}_{ab}^{Pol}  \ket{\Phi^{+}}_{ab}^{Ph}$ \\
phase error: 1} \\
\addlinespace[1.5ex]
\multirow{1}{*}{$D_{2H}, D_{1V}$} & \multirow{1}{*}{Odd}
& \makecell{$\ket{\Psi^{-}}_{ab}^{Pol}  \ket{\Psi^{+}}_{ab}^{Ph}$ \\
correct BSM  result}
& \makecell{$\ket{\Psi^{-}}_{ab}^{Ph}  \ket{\Psi^{-}}_{ab}^{Ph}$\\
phase error: 1} \\
\addlinespace[1.5ex]
\multirow{1}{*}{$D_{2H}, D_{1V}$} & \multirow{1}{*}{Even}
& \makecell{$\ket{\Psi^{+}}_{ab}^{Pol}
\ket{\Psi^{-}}_{ab}^{Ph}$ \\
phase error: 2}
& \makecell{$\ket{\Psi^{+}}_{ab}^{Pol}  \ket{\Psi^{+}}_{ab}^{Ph}$ \\
phase error: 1} \\
\addlinespace[1ex]
\doublebottomrule
\end{tabular}
\label{table4}
\end{table}

In the same way, the items corresponding to $D_{2H}$ and $D_{1V}$ responding can be written as
\begin{equation}
\begin{aligned}
&|00\rangle_a|11\rangle_{b}|\sqrt{\mu^\prime}\rangle_{H_2}|\sqrt{\mu^\prime}\rangle_{V_1}
+|01\rangle_a|10\rangle_{b}|-\sqrt{\mu^\prime}\rangle_{H_2}|-\sqrt{\mu^\prime}\rangle_{V_1}\\
&+|10\rangle_a|01\rangle_{b}|\sqrt{\mu^\prime}\rangle_{H_2}|-\sqrt{\mu^\prime}\rangle_{V_1}
+|11\rangle_a|00\rangle_{b}|-\sqrt{\mu^\prime}\rangle_{H_2}|\sqrt{\mu^\prime}\rangle_{V_1},
\end{aligned}
\end{equation}
the bit's correlation corresponding  can be written as
\begin{equation}
\begin{aligned}\label{event3}
&\left|01\right\rangle_{ab}^{Pol}{[\left|01\right\rangle}_{ab}^{Ph}+{(-1)}^{(m+n)}\left|10\right\rangle_{ab}^{Ph}]
+{\left|10\right\rangle_{ab}^{Pol}[(-1)}^n\left|01\right\rangle_{ab}^{Ph})+{(-1)}^m\left|10\right\rangle_{ab}^{Ph}].
\end{aligned}
\end{equation}
where $m$ and $n$ denote the photon number of the coherent states arriving at $D_{2H}$ and $D_{1V}$, respectively. From Eq. (\ref{event3}), Charlie can obtain the key bits as 11. We also define the correct BSM results corresponding to both $m$, $n$ being odd numbers.

We define the phase error rate $P_r^{H_1V_2(H_2V_1)}(e,e)$, $P_r^{H_1V_2(H_2V_1)}(o,e)$, $P_r^{H_1V_2(H_2V_1)}(e,o)$  corresponding to the phase error rates of the three different odd-even combinations of $m$ and $n$.

\section{Analysis of detector responses considering channel loss and dark counts}
Considering more general and practical scenarios, the effects of channel loss and dark counts would influence the average photon number and the detector response.
We take the quantum state in Eq. \eqref{eq.6} as an example where both parties encode their WCPs with bits $00$. After considering channel loss and dark counts, the state in Eq. \eqref{eq.6} is transformed to
$|0000\rangle_{ab}|\sqrt{\eta\mu}\rangle_{D_{1H}}|\sqrt{\eta\mu}\rangle_{D_{1V}}$, where $\eta$ is the total transmission efficiency $\eta=\eta_{t}\eta_d$ ($\eta_d$ is the detection efficiency).

In Event$_1$, the response probability of $D_{1H}$ is $P_r(D_{1H})=1-(1-P_d)P_{I_{H_1}}(0)$, where $P_d$ is the dark count rate, $P_{I_{H_1}}(0)=e^{-I_{H_1}}$ denotes the probability of the detector not responding caused by the vacuum state ($I_{H_1}={(\sqrt{\eta\mu})}^2=\eta\mu$ is the intensity of the input pulse of $D_{1H}$).
In this way, the probability corresponding to only $D_{1H}$ responding caused by an even number of photons excluding the vacuum state and dark counts is
\begin{equation}
\label{eq24}
\begin{aligned}
&P_r^{H_1}(e)=(1-P_r(D_{2H}))(1-P_r(D_{1V}))(1-P_r(D_{2V}))\sum_{i>0}{P_r^{H_1}\left(2i\right)}\\
& \quad=(1-P_d)^3e^{(-I_{H_2}-I_{V_1}-I_{V_2})}\sum_{i>0}{P_r^{H_1}\left(2i\right)}\\
& \quad=[(1-P_d)^3e^{(-I_{H_2}-I_{V_1}-I_{V_2})}]\sum_{i>0} e^{-I_{H_1}}\frac{I_{H_1}}{(2i)!}-e^{-I_{H_1}}\left(1-P_d\right)\\
& \quad=(1-P_d)^3e^{(-I_{H_1}-I_{H_2}-I_{V_1}-I_{V_2})}[cosh(I_{H_1})-1+P_d].
\end{aligned}
\end{equation}

In Event$_2$, we consider the probability that both $D_{1H}$ and $D_{1V}$ respond caused by an even number of photons excluding the vacuum state and dark counts as
\begin{equation}
\begin{aligned}
P_r^{H_1V_1}(e,e)=(1-P_d)^2e^{(-I_{H_1}-I_{H_2}-I_{V_1}-I_{V_2})}[cosh(I_{H_1})-1+P_d][cosh(I_{V_1})-1+P_d].
\end{aligned}
\end{equation}
Similarly, we can also calculate $P_r^{H_1V_1}(e,o)$ and $P_r^{H_1V_1}(o,e)$ as
\begin{equation}
\begin{aligned}
P_r^{H_1V_1}(o,e)=(1-P_d)^2e^{(-I_{H_1}-I_{H_2}-I_{V_1}-I_{V_2})}sinh(I_{H_1})[cosh(I_{V_1})-1+P_d].
\end{aligned}
\end{equation}
\begin{equation}
\begin{aligned}
P_r^{H_1V_1}(e,o)=(1-P_d)^2e^{(-I_{H_1}-I_{H_2}-I_{V_1}-I_{V_2})}[cosh(I_{H_1})-1+P_d]sinh(I_{V_1}).
\end{aligned}
\end{equation}

\section{The formula derivation of $Q_{Event_i}$, $E_{Event_i}^{bit}$ and $E_{Event_i}^{ph}$ ($i=1,2,3$)}

The key rate of our dual-DOF QSS protocol can be calculated as \cite{INI_qkd}
\begin{equation}\label{R}
R=\sum_{i=1}^{3}{Q_{Event_i}[1-I_E-H\left(E_{Event_i}^{ph}\right)-fH\left(E_{Event_i}^{bit}\right)]}.
\end{equation}
Here, $Q_{Event_i}$ denotes the overall gain in Event$_i$ $(i \in \{1, 2, 3\})$, $f$ denotes the error-correction efficiency. $E_{Event_i}^{ph}$ and $E_{Event_i}^{bit}$ denote the phase-error rate and bit-error rate of Charlie's BSM results in Event$_i$ $(i \in \{1, 2, 3\})$, respectively and $H(x) = - xlog_2x - (1-x) log_2(1 - x)$ is the binary Shannon entropy function.

In the parameter calculation, we focus on specific cases where Alice's and Bob's phase encoded bits are (0, 0), and polarization encoded bits are (0, 0) and (0, 1) to exemplify the general scenario due to the symmetry \cite{INI_qkd}. The wrong detector responses caused by the bit-flip error and the correct detector responses under these two cases are shown in Tab. \ref{table1} as
 \begin{table}[htbp]
\centering
\caption{The correct detector responses ($"\checkmark"$) and wrong detector responses ($"\times"$) caused by the bit-flip error in $(0000)$ and $(0001)$ scenarios. $"\circ"$ means that the encoded bits has no correlation from the detector response. }
\setlength{\arrayrulewidth}{0.6pt}
\renewcommand{\arraystretch}{1.3} 
\begin{tabular}{c|cc|cc|cc}
	\hline
	\hline
	\multirow{2}{*}{\makecell{bit correlation\\$(\kappa_{a}^{Ph} \kappa_{b}^{Ph} \kappa_{a}^{Pol} \kappa_{b}^{Pol})$}}
	& \multicolumn{2}{c|}{Event$_1$} & \multicolumn{2}{c|}{Event$_2$} & \multicolumn{2}{c}{Event$_3$} \\
	\cline{2-7}
	& \scalebox{0.8}{$D_{1H}$} & \scalebox{0.8}{$D_{2H}$} & \scalebox{0.8}{$D_{1H},D_{1V}$} & \scalebox{0.8}{$D_{2H},D_{2V}$} & \scalebox{0.8}{$D_{1H},D_{2V}$} & \scalebox{0.8}{$D_{2H},D_{1V}$} \\
	\hline
	$Pol(0000)$ & \scalebox{1.6}{$\circ$} & \scalebox{1.6}{$\circ$} & $\checkmark$ & $\checkmark$ & $\times$ & $\times$ \\
	$Ph(0000)$ & $\checkmark$ & $\times$ & $\checkmark$ & $\times$ & $\checkmark$ & $\times$ \\
	$Pol(0001)$ & \scalebox{1.6}{$\circ$} & \scalebox{1.6}{$\circ$} & $\times$ & $\times$ & $\checkmark$ & $\checkmark$ \\
	$Ph(0001)$ & $\checkmark$ & $\times$ & $\checkmark$ & $\times$ & $\checkmark$ & $\times$ \\
	\hline
	\hline
\end{tabular}
\label{table1}
\end{table}

Due to the symmetry of the system, the probabilities of the combinations ``$++$'' and ``$--$'' are equivalent, and those of the combinations ``$+-$'' and ``$-+$'' are equivalent. Therefore, it suffices to consider only the two representative polarization combinations ``$++$'' and ``$+-$''(thus each with probability 1/2). For each Event$_i$, the gain $Q_{Event_i}$ is determined by the sum of detector responses as defined (corresponding to the $\checkmark$ and $\times$ entries in Tab. \ref{table1}). The wrong detector responses caused by the bit-flip error are marked with $\times$.The phase error rate of the BSM in each event is detailed in Sec. \ref{sec222}. Finally, the bit and phase error rates for each scenario are summarized in Tab. $\mathrm{III}$ in the maintext.

In Event$_1$, we can calculate the total gain $Q_{Event_1}$, the bit error rate of the BSM results $E_{Event_1}^{bit}$ and the phase error rate of the BSM results $E_{Event_1}^{ph}$ in the $X_{dual}$ basis as
\begin{equation}
\begin{aligned}\label{event1n}
 Q_{Event_1}&=P_r(D_{1H})+P_r(D_{2H})\\
 &=\frac{1}{2}[P_r(D_{1H}|++)+P_r(D_{1H}|+-)+P_r(D_{2H}|++)+P_r(D_{2H}|+-)],\\
E_{Event_1}^{bit}&=\frac{1}{2Q_{Event_1}}[P_r(D_{2H}|++)+P_r(D_{2H}|+-)],\\
E_{Event_1}^{ph}&=\frac{P_r^{H_1}(e|++)+P_r^{H_1}(e|+-)}{2Q_{Event_1}}.
\end{aligned}\end{equation}
 Taking Eq. (\ref{eq24}) as an example, if the polarization bits are ``$++$", $P_r^{H_1}(e)$ can be transformed into $P_r^{H_1}(e|++)=(1-P_d)^3e^{(-I_{H_1}^{++}-I_{H_2}^{++}-I_{V_1}^{++}-I_{V_2}^{++})}[cosh(I_{H_1}^{++})-1+P_d]$.

Similarly, referring to Tab. \ref{table3} and Tab. \ref{table4}, we can calculate $Q_{Event_2}$, $E_{Event_2}^{bit}$, $E_{Event_2}^{ph}$ and $Q_{Event_3}$, $E_{Event_3}^{bit}$, $E_{Event_3}^{ph}$ in the $X_{dual}$ basis as
\begin{equation}
\begin{aligned}\label{event2n}
Q_{Event_2}&=P_r(D_{1H}D_{1V}|++)+P_r(D_{1H}D_{1V}|+-)+P_r(D_{2H}D_{2V}|++)+P_r(D_{2H}D_{2V}|+-),\\
E_{Event_2}^{bit}&=\frac{1}{2Q_{Event_2}}[P_r(D_{1H}D_{1V}|+-)+P_r(D_{2H}D_{2V}|++)+2P_r(D_{2H}D_{2V}|+-)],\\
E_{Event_2}^{ph}&=\frac{2P_r^{H_1V_1}[(o,e)|++]+P_r^{H_1V_1}[(e,o)|++]+P_r^{H_1V_1}[(e,e)|++]}{2Q_{Event_2}}\\
& +\frac{P_r^{H_1V_1}[(e,o)|+-]+P_r^{H_1V_1}[(o,e)|+-]}{2Q_{Event_2}}+\frac{P_r^{H_2V_2}[(e,e)|++]+P_r^{H_2V_2}[(o,e)|++]}{2Q_{Event_2}},
\end{aligned}
\end{equation}

\begin{equation}
\begin{aligned}\label{event3n}
Q_{Event_3}=&P_r(D_{1H}D_{2V}|++)+P_r(D_{1H}D_{2V}|+-)
+P_r(D_{1V}D_{2H}|++)+P_r(D_{1V}D_{2H}|+-),\\
E_{Event_3}^{bit}=&\frac{1}{2Q_{Event_3}}[P_r(D_{1H}D_{2V}|++)+2P_r(D_{1V}D_{2H}|++)+P_r(D_{1V}D_{2H}|+-)],\\
E_{Event_3}^{ph}=&\frac{P_r^{H_1V_2}[(o,e)|++]+P_r^{H_1V_2}[(e,o)|++]}{2Q_{Event_3}}\\
+&\frac{P_r^{H_1V_2}[(e,o)|+-]+2P_r^{H_1V_2}[(o,e)|+-]+P_r^{H_1V_2}[(e,e)|+-]}{2Q_{Event_3}}\\
+&\frac{P_r^{H_2V_1}[(e,e)|++]+P_r^{H_2V_1}[(o,e)|++]}{2Q_{Event_3}}.
\end{aligned}
\end{equation}

\begin{figure}[htbp]
\centering
\includegraphics[scale=0.5]{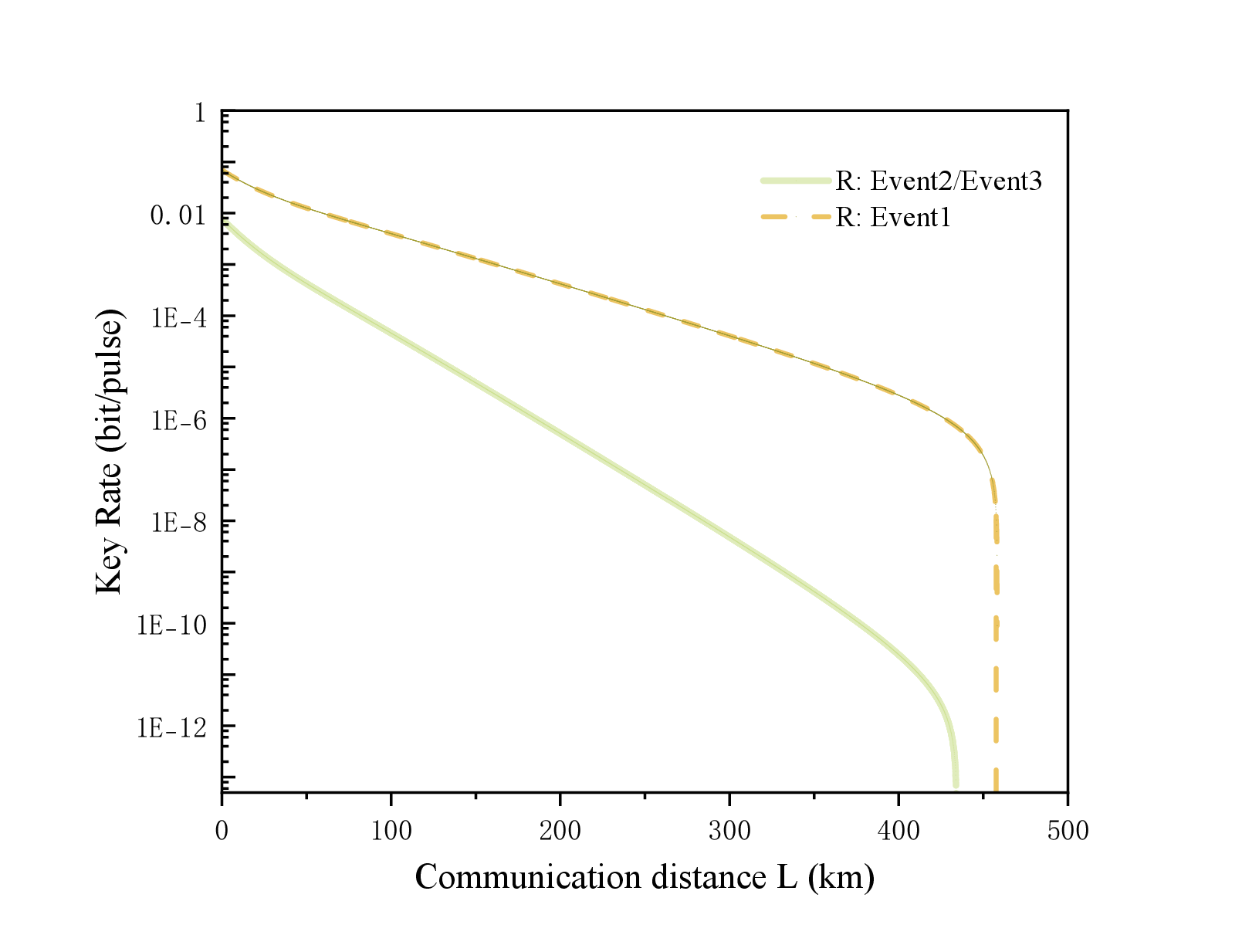}
\caption{The key rates in Event$_1$, Event$_2$, Event$_3$ altered with the communication distance under $\mu=0.84$. Here, we set $\eta_d=0.145$, the attenuation coefficient of the fiber $\alpha=0.2$ dm/km, the dark count rate $P_d=8 \times 10^{-8}$, and the error correction efficiency $f= 1.15$. }
\label{FIG.1.0_supplement}
\end{figure}

Finally, by taking Eq. (\ref{event1n})-Eq. (\ref{event3n}) into Eq. (\ref{R}), we can obtain the key rate as a function of transmission distance in each event under $\mu=0.84$. Here, we set $\eta_d=0.145$, the attenuation coefficient of the fiber $\alpha=0.2$ dm/km, the dark count rate $P_d=8 \times 10^{-8}$, and the error correction efficiency $f= 1.15$. As shown in the Fig. \ref{FIG.1.0_supplement}, the maximum transmission distance in Event$_1$ reaches about $458.3 km$, while that in Event$_2$ (Event$_3$) reaches about $434.3 km$. The threshold of the QBER in Event$_1$ is 2.39\%, while that in Event$_2$ (Event$_3$) is about 2.08\%.